\documentclass[12pt,nohyper]{JHEP3}         
\usepackage{amssymb,amsfonts}
\font\mybb=msbm10 at 12pt
\def\bb#1{\hbox{\mybb#1}}

\def\bR {\bb{R}}
\def\bE {\bb{E}}
\def\bI {\bb{I}}

\newcommand{\be}{\begin{equation}}
\newcommand{\ee}{\end{equation}}
\newcommand{\bea}{\begin{eqnarray}}
\newcommand{\eea}{\end{eqnarray}}

\newcommand{\nn}{\nonumber \\}

\def\6{\partial}
\def\7{\tilde}
\def\8{\widehat}

\def\G11{\Gamma_{11} }

\title{SDiff Gauge Theory and the M2 Condensate}

\author{Igor A. Bandos \\
Ikerbasque, Basque Science Foundation and \\ Department of Theoretical Physics and History of Science, \\
 The University of the Basque Country, \\ P.O. Box 644,
48080 Bilbao, Spain \\  and \\ ITP, NSC KIPT, 61183, Kharkov,
Ukraine}

\author{Paul K. Townsend\\
Department of Applied Mathematics and Theoretical Physics \\
Centre for Mathematical Sciences, University of Cambridge\\
Wilberforce Road, Cambridge, CB3 0WA, UK.}

\abstract{We develop a general formalism for the construction, in
$D$-dimensional Minkowski space,  of  gauge theories  for which the
gauge  group is the infinite-dimensional group SDiff$_n$ of
volume-preserving diffeomorphisms  of some  closed $n$-dimensional
manifold.   We then focus on the $D=3$ SDiff$_3$ superconformal  gauge theory
describing  a condensate of M2-branes; in particular, we derive its ${\cal N}=8$ superfield
equations from  a pure-spinor superspace action, and we  describe its relationship
to the  $D=3$ SDiff$_2$ super-Yang-Mills theory describing a condensate of
D2-branes. }

\preprint {DAMTP-2008-65, arXiv:0808.1583}

\begin{document}
\section{Introduction}
\setcounter{equation}{0}

The M2-branes of M-theory may have boundaries on an M5-brane because
the M2-charge can be taken up by the 2-form gauge potential on the
M5-brane worldvolume \cite{Strominger:1995ac,Townsend:1995af}.
Following the determination of the M5-brane equations of motion
\cite{Howe:1996yn} and the construction of  its action
\cite{Bandos:1997ui},  it  was verified that there exists a
`soliton-type' solution with this  interpretation
\cite{Howe:1997ue}. This possibility can also be understood from the
M2-brane perspective in terms of  its  superalgebra
\cite{Bergshoeff:1997bh}, and is realizable in terms of an open
membrane subject to appropriate boundary conditions
\cite{Chu:1998pb} but not, for a single M2-brane,  as a
`soliton-type' solution of the M2-brane equations of motion. This is
hardly surprising given the disparity in dimension but one may
imagine that multiple M2-branes could expand to generate the
required extra dimensions as a `fuzzy' 3-sphere, and an equation
that might describe such a configuration was proposed  by Basu and
Harvey \cite{Basu:2004ed}. This equation  led Bagger and Lambert  to
propose \cite{Bagger:2007jr}, as a low-energy limit of coincident
planar  M2-branes,  a novel class of 3-dimensional  maximally
supersymmetric gauge theories  based on Filippov 3-algebras, rather
than Lie algebras;  a similar framework was developed by  Gustavsson
\cite{Gustavsson:2007vu}. Such gauge theories have the  $OSp(8|4)$
superconformal symmetry expected of an action for multiple M2-branes
in a low-energy  limit \cite{Bandres:2008vf},  and they admit the
Basu-Harvey equation as  a `BPS'  equation.

Explicit  realizations of the  Bagger-Lambert-Gustavsson (BLG)
theory  arise from specific  Filippov  3-algebras.  A particular
4-dimensional example, ${\cal A}_4$, was considered by Bagger and
Lambert  \cite{Bagger:2007jr} but  the corresponding BLG model has
since been shown \cite{Lambert:2008et,Distler:2008mk} to describe
the  dynamics of two  M2-branes on an orbifold rather than flat
space.  This model is also disappointing in one other respect: it is
equivalent to a `standard' Chern-Simons (CS) theory for gauge group
$SU(2)\times SU(2)$ coupled to  ${\cal N}=8$ matter multiplets in
the $({\bf 2},{\bf 2})$ representation \cite{VanRaamsdonk:2008ft},
so the novel algebraic structure of the general construction plays
no essential role in this example. Furthermore, all other
finite-dimensional Filippov `metric'  3-algebras (those with
positive definite algebra-compatible metric) are direct sums of
${\cal A}_4$ and trivial one-dimensional 3-algebras
\cite{Papadopoulos:2008sk,Gauntlett:2008uf}, so the nature of the
action describing the low-energy dynamics of an arbitrary finite
number $N$ of coincident planar M2-branes remains an unsolved
problem, although there is no shortage of proposals.  We will return
to this point at the conclusion of this paper;  for most purposes
here it is sufficient  that there are clear candidates for the
$N\to\infty$ limit,  which one can view as describing possible
`condensates' of coincident planar M2-branes.  These are the BLG
theories in which the Filippov 3-algebra is realized by the
Nambu-bracket \cite{Nambu:1973qe} of functions defined on some
3-manifold $M_3$;  the choice $M_3=S^3$ then leads to a
version of the Basu-Harvey equation in which the fuzzy 3-sphere
becomes a classical 3-sphere \cite{Bagger:2007vi}.

Recall that the Nambu
$n$-bracket for  $n$ functions $(\phi^1,\dots, \phi^n)$ on a closed
$n$-dimensional manifold $M_n$ with coordinates $\sigma^i$
($i=1,\dots,n$) is
 \be \left\{ \phi^1,\dots, \phi^n\right\} = e^{-1} \varepsilon^{i_1\dots i_n}
\partial_{i_1} \phi^1 \cdots \partial_{i_n}\phi^n\, ,
\ee
where $\varepsilon$ is the invariant antisymmetric tensor
density on $M_n$. We choose to define this bracket  as a scalar on
$M_n$ by dividing by  some fixed scalar density $e$ on $M_n$. The
space of functions on $M_n$ can then be viewed as an
infinite-dimensional `$n$-algebra'.  This algebra obeys a
`fundamental'  identity that can be expressed simply in terms of two
{\it anticommuting} (`ghost') functions  $(B,C)$ on $M_n$:
 \be \{B, \dots,B,  \{C, \dots,
C\}\} = n \{  \{B,\dots,B,C\},C,\dots ,C\}\, .
\ee
Abstractly,  any $n$-algebra
defined by an $n$-linear antisymmetric product that obeys the fundamental identity is a
Filippov $n$-algebra. The $n$-algebra of the space of functions on $M_n$ with respect
to the Nambu $n$-bracket is therefore an infinite-dimensional Filippov $n$-algebra.

One should not think of  the density $e$ on $M_n$ as derived from a
metric on  $M_n$ because no metric will be used in our
constructions, but one may choose $e$ to coincide with $\sqrt{g}$
for some `fiducial' metric $g$ that one could introduce for this
purpose. For example, if $M_n \cong S^n$ then one may choose $e=
\sqrt{g}$ where $g$ is the $SO(n+1)$-invariant metric on the unit
$n$-sphere. This choice facilitates the identification of  the
finite-dimensional sub-algebra  that exists when $M_n \cong S^n$.
Consider $(n+1)$ functions $X_a$ ($a=1,\dots,n+1$) subject to the
constraint \be\label{unitS=1} \sum_{a=1}^{n+1} X_a^2 = 1\, . \ee
Given that $e$ has been chosen as specified above, then
\be\label{unitS} \left\{X_{a_1},\dots X_{a_n}\right\} =
\epsilon^{a_1\dots a_n a_{n+1}}X_{a_{n+1}} \, , \ee which shows that
the $X_a$ span an $(n+1)$-dimensional subalgebra:   ${\cal A}_n$.
For $n=2$ the Nambu bracket is a Poisson bracket and we therefore
have a realization of the Lie algebra $su(2)$ by functions on $S^2$,
so ${\cal A}_3= su(2)$.  For $n=3$  we have a realization of the
four-dimensional Filippov 3-algebra ${\cal A}_4$ by functions on
$S^3$.

As suggested in \cite{Bagger:2007vi} and
shown in  \cite{Ho:2008nn,Ho:2008ve,Bandos:2008fr}, the Nambu
bracket realization  of the BLG theory is  an `exotic'  gauge theory
for the group SDiff($S^3$) of volume-preserving diffeomorphisms  of
the 3-sphere. A rather explicit  discussion of this group is given in
\cite{Dowker:1990iy};  other 3-manifolds $M_3$ yield slightly different
theories; we return to this point in the final section but otherwise pass over it,
using the notation SDiff$_3$  for the group of volume-preserving diffeomorphisms
of any closed 3-manifold $M_3$.  We say that  SDiff$_3$ gauge theories are
`exotic'   because they
cannot be obtained from an `abstract' YM theory,  whereas this is possible for SDiff$_2$
gauge theories; we elaborate on this this point  later.  Since the fields of an SDiff$_3$ gauge
theory also depend on the three coordinates of $M_3$, the Nambu bracket realization of
the BLG theory  is effectively a 6-dimensional theory.  It has been
suggested that this is a version of the M5-brane action
\cite{Ho:2008nn,Ho:2008ve}, although the most straightforward
way to extract an SDiff$_3$ gauge theory from the {\it standard} M5-brane action
leads to the Carrollian limit of the BLG theory  \cite{Bandos:2008fr}.

The main aim of this paper  is to put the Nambu-bracket realization of the BLG theory into a larger context by developing  further the general principles of SDiff gauge theory.   It is well-known that SDiff$_2$ gauge theories may loosely be considered as $N\to\infty$ limits of $SU(N)$ gauge theories in which the matrix commutator becomes the Poisson bracket of functions on a
2-manifold \cite{LCGM2-SU(inf)}, the 2-sphere being the simplest case.  Such theories first arose  from light-cone gauge-fixing of  a relativistic membrane, and the application to the M2-brane yields a maximally supersymmetric gauge  mechanics  model  in which the gauge group is the infinite-dimensional group of area-preserving diffeomorphisms of the membrane. In the case of a spherical membrane, there is a sequence of truncations of the group of area-preserving diffeomorphisms to $SU(N)$ that reduces the membrane action to the action for a maximally-supersymmetric $SU(N)$ gauge mechanics model \cite{de Wit:1988ig}; this truncation is one in which the classical 2-sphere is replaced by a fuzzy sphere \cite{Madore:1991bw}. The truncated model can be interpreted as describing the dynamics of multiple D0-branes \cite{Townsend:1995af}, and is the basis of the M(atrix) model formulation of  M-theory \cite{Banks:1996vh}.

In the context of gauge mechanics models,  which we may view as
examples of $D$-dimensional gauge theories for $D=1$,  there exist
SDiff$_n$ gauge theories for any $n=p$ obtained by the light-cone gauge-fixing
of the action for a relativistic $p$-brane \cite{Bergshoeff:1988hw} (although supersymmetry
constrains $p$ and hence $n$).  What we are interested in
this paper is how SDiff$_n$ gauge theories may be constructed for
$D>1$. The answer to this question for $n=2$ is known. Because
SDiff$_2$ gauge theories are just standard, albeit
infinite-dimensional, Yang-Mills theories, any Yang-Mills theory
that can be constructed for all $SU(N)$ can also be constructed for
SDiff$_2$ \cite{Floratos:1988mh}.  For example, one may choose the
gauge group for the $D=4$ ${\cal N}=4$ super-Yang-Mills theory to be
SDiff($S^2$), in which case we have a 6-dimensional theory. It is
possible that this is related to the M5-brane in the much the
same way as the Nambu-bracket realization of the BLG theory, but we shall
not investigate this possibility here.  Instead, we focus on possibilities
for SDiff$_n$ gauge theories with $n>2$.

It appears that there are no useful possibilities  for $n\ge 4$
because of the difficulty in constructing a kinetic term for the
gauge potential without a metric on $M_n$. For this reason, we focus
on the $n=3$ case. Remarkably, SDiff$_3$ gauge theories may be
constructed for any spacetime dimension $D$ in close analogy to
Yang-Mills theory, although  these theories are still `exotic' in
the sense explained above. However, they are unlikely to be of any
physical relevance because their energy density is  not positive
definite. For $D=3$ there is another option: one may construct a
Chern-Simons-type  term.  This leads to a new class of
(super)conformal $D=3$ gauge theories, which we focus on in
this paper. The Nambu bracket realization of the BLG theory is the
maximally-supersymmetric SDiff$_3$ gauge theory of this
type, and  we re-construct it from  our formalism,  presenting
simple proofs of  both its ${\cal N}=8$ supersymmetry and its
superconformal invariance.  Although there is no free field limit of the
BLG action, we show that one can take a free-field limit of  the
equations of motion, in which case one arrives at a theory for an
infinite number of non-interacting ${\cal N}=8$ scalar supermultiplets
related by a  {\it rigid} SDiff$_3$ symmetry.

As we are attempting to put the BLG model into a more general
context, we consider the general construction of superconformal
SDiff$_3$  gauge theories in terms of ${\cal N}=1$
superfields\footnote{An ${\cal N}=1$ formulation of the abstract  BLG theory
was proposed previously in \cite{Mauri:2008ai} but the Nambu
bracket realization  was not spelled out there.}. Obviously, any
(Minkowski space) SDiff$_3$ gauge  theory with ${\cal N}>1$
supersymmetry can be written in terms of ${\cal N}=1$ superfields,
although the extended supersymmetry will not then be manifest.
To make the ${\cal N}=8$ supersymmetry of the BLG theory manifest, one
needs a formulation of it in terms of ${\cal N}=8$ superfields.  After
the  original version of this paper appeared on the archives, two
distinct proposals were made for an  ${\cal N}=8$ superfield
formulation: one an off-shell  formulation of the abstract BLG theory
\cite{Cederwall:2008vd,Cederwall:2008xu} using a `pure-spinor superspace',
the other an on-shell  ${\cal N}=8$ superfield formulation of the Nambu bracket
realization of the BLG theory \cite {Bandos:2008df}. Here we review the latter
approach, with some simplifications, and we explain how the former approach
extends to the Nambu-bracket realization of the BLG theory.

The low-energy dynamics of $N$ coincident (or nearly-coincident) parallel planar
D2-branes is an  ${\cal N}=8$ supersymmetric  $D=3$ gauge theory with gauge group
$SU(N)$. As explained above, $SU(N)$ can be viewed as a finite-dimensional approximation
to  SDiff$_2$ (at least when $M_2=S^2$). It follows that the ${\cal N}=8$ supersymmetric
Yang-Mills theory with gauge group SDiff$_2$  may be interpreted as the field theory describing the low-energy dynamics of a D2-condensate, in much the same sense as the BLG theory describes
an M2-condensate. In fact, we expect the renormalization group flow of a model for the D2-condensate to yield, in the infra-red limit, a model for the M2-condensate because this
limit  decompactifies  IIA superstring theory to M-theory. Conversely, one might expect
an $S^1$-compactification of a model for the M2-condensate to yield a model for the
D2-condensate. Here we show that the ${\cal N}=8$ supersymmetric   $D=3$ SDiff$_2$ Yang-Mills theory is indeed an $S^1$-compactification of the SDiff$_3$ BLG theory, in a sense that we make precise. We also show that this model is an $S^1$-compactification, in a different sense, of the
 ${\cal N}=4$ supersymmetric   $D=4$ SDiff$_2$ Yang-Mills theory mentioned above.

\section{SDiff gauge theory}
\label{sec:SDiffGI}
\setcounter{equation}{0}

Let $M_n$ be a closed $n$-dimensional real manifold that is compact
with respect to some (non-dynamical) scalar density  $e$ in local
coordinates $\sigma^i$ ($i=1,\dots,n$). In new local
coordinates $\sigma^i + \xi^i(\sigma)$, for infinitesimal vector
field $\xi$, the  scalar density becomes $e - \partial_i(e\xi^i)$,
so the total volume is unchanged (as expected since this cannot
depend on the choice of coordinate atlas for $M_n$) but the local
volume density changes unless we impose the constraint
\be\label{con-xi}
\partial_i \left(e\xi^i\right) =0\, .
\ee
The space of vector fields on $M_n$ satisfying this constraint is a subalgebra of the algebra of all vector fields with respect to the Lie bracket of vector fields. It is the Lie algebra of the group
SDiff($M_n$) of `volume-preserving' diffeomorphisms of $M_n$, which we abbreviate to SDiff$_n$.

We are concerned here with field theories in $D$-dimensional
Minkowski spacetime, with cartesian coordinates $x^\mu$, and `mostly
plus' metric $\eta_{\mu\nu}$. Consider a scalar field $\phi$ that is
also a scalar on $M_n$; it can be expanded in $M_n$-harmonics so
$\phi$ contains an infinity of Minkowski scalar fields, which
transform among themselves under the infinite-dimensional group
SDiff$_n$. The infinitesimal SDiff$_n$ transformation of $\phi$ is
\be\label{varScal} \delta_\xi \phi = - \xi^i\partial_i \phi\, . \ee
More generally, for any Minkowski-field $T$ that is also a tensor on
$M_n$, the infinitesimal SDiff$_n$ transformation is
\be\label{vSDiffT} \delta_\xi T= -{\cal L}_\xi T\, , \ee where
${\cal L}_\xi$ is the Lie derivative with respect to $\xi$.
Besides (\ref{varScal}),  other  important special cases are
\be\label{vSDiffS} \delta_\xi v^i=
-\xi^j\partial_j v^i + v^j\partial_j \xi^i\,\,  , \qquad \delta_\xi
\omega_i= -\xi^j\partial_j \omega_i - (\partial_i \xi^j)\,
\omega_j\, . \ee for  vector $v^i$ and one-form $\omega_i$ on $M_n$.

It is not difficult to construct Minkowski-space field theories that
have a {\it rigid} SDiff$_n$ invariance. For example, the Lagrangian
density.
 \be\label{rigid} {\cal L} =  \oint \!
d^n\sigma \, e\,\left[ -\frac{1}{2} \eta^{\mu\nu}  \partial_\mu
\phi\, \partial_\nu\phi -V(\phi)\right]
 \ee
is SDiff$_n$-invariant, for any potential  function $V$,  as long as
the $M_n$-vector parameter  $\xi$ is independent of the
Minkowski space coordinates. This is an interacting Lagrangian
density for the infinite number of Minkowski scalar fields contained
in the $M_n$-harmonic expansion of $\phi$. However, we are
interested in constructing SDiff$_n$ {\it gauge} theories for which
the SDiff$_n$ invariance is local, in the sense that $\xi$
is allowed to be an arbitrary Minkowski scalar in addition to being
a divergence-free $M_n$-vector field. This will require new
ingredients, as we explain next.

\subsection{Local SDiff$_n$ invariance}

The Minkowski spacetime  derivative   $dT = dx^\mu\partial_\mu T$ is
again a tensor on $M_n$ (of the same type) as long as $\xi$ is assumed to be
independent  of  the Minkowski spacetime coordinates,  but if we
insist on {\it local}  SDiff$_n$ invariance  then we need to use the covariant exterior derivative
\be
{\cal D}= d + {\cal L}_s\, , \qquad s= dx^\mu s_\mu{}^i\, \partial_i\, ,
\ee
where the one-form-valued $M_n$-vector field $s$ satisfies the constraint
\be\label{con-s}
\partial_i\left(es^i \right) \equiv 0\, .
\ee
One may verify, for $M_n$-tensor $T$, that
 \be
 \delta_\xi \left({\cal D}T\right) = - {\cal L}_\xi \left({\cal D}T\right)\, ,
 \ee
 provided that we assign to $s$  the  SDiff$_n$ gauge transformation
 \be
 \delta_\xi s = d\xi- [\xi,s] \, ,
 \ee
where the bracket [,] indicates a commutator of vector fields on
$M_n$.   In particular, for $M_n$-scalar $\phi$,
\be
{\cal D}\phi=d\phi + s^i\partial_i\phi\, , \qquad
\delta_\xi ({\cal D}\phi) =- \xi^i\partial_i ({\cal D}\phi)\, .
\ee
Note that the constraint (\ref{con-s}) is SDiff$_n$ invariant as  a consequence
of  (\ref{con-xi}).

This  formalism may be extended to tensor densities on $M_n$. In particular, the
SDiff gauge transformation of the scalar density $e$ is zero because
of the constraint (\ref{con-xi}). We assume that $e$ is independent of the Minkowski
coordinates; i.e.
\be\label{de0}
de =0 \qquad \left(\Leftrightarrow \ \partial_\mu e =0\right)\, .
\ee
As a consequence, one may show that\footnote{For example, both sides vanish when $T=1$, the left hand side because $de=0$ and the right hand side by the definition of ${\cal D} T$.}
\be\label{DeT}
{\cal D}\left(eT\right) = e{\cal D}T\, .
\ee

As for Yang-Mills gauge theories, we may define the covariant 2-form field-strength of  $s$
as\footnote{We use the convention in which $d$ acts  `from the left'.}
\be
F= ds + \frac{1}{2}\left[s,s\right]\, .
\ee
This has the SDiff gauge transformation
\be
\delta_\xi  F = -[\xi,F]\, ,
\ee
and it satisfies the `Bianchi' identity ${\cal D}F\equiv 0$, i.e.
\be
dF  + \left [s, F\right] \equiv 0 \, .
\ee
We may write $F= F^i \partial_i$, where
\be\label{fieldstrength}
F^i = ds^i + s^j\partial_j s^i\, .
\ee
This satisfies the additional identity
\be\label{divfree}
\partial_i \left(e F^i\right)\equiv 0\, .
\ee

\subsection{Pre-gauge invariance}

The constraints (\ref{con-xi}) and (\ref{con-s}) may be solved,
locally, by writing \be e\xi^i = \varepsilon^{ijk_1  \dots
k_{n-2}}\partial_j \omega_{k_1\dots k_{n-2}}\, , \qquad es^i =
\varepsilon^{ijk_1  \dots k_{n-2}}\partial_j A_{k_1\dots k_{n-2}}\,
, \ee where the   $(n-2)$-form $\omega$ (on $M_n$) is an
unconstrained parameter, and $A$ is an $(n-2)$-form pre-potential on
$M_n$ (in addition to being a 1-form on the $D$-dimensional
Minkowski spacetime); its  SDiff$_n$ transformation
 is\footnote{In our convention,  square brackets indicate
antisymmetrization of the indices enclosed with `strength
one' (so that the brackets may be simply omitted on contraction of
all antisymmetrized indices with some other antisymmetric tensor).}
\be \delta_\xi A_{i_1\dots i_{n-2}} = d\omega_{i_1\dots i_{n-2}}
-\xi^j\partial_j A_{i_1\dots i_{n-2}} -
\left(n-2\right)\partial_{[i_1}\xi^j A_{|j|i_2\dots i_{n-2}]}\, .
 \ee
In addition, for $n \ge3$, we have the abelian  pre-gauge
transformation\footnote{This holds also for $n=2$ if  we view the
pre-gauge transformation as a shift of $A$ by  a closed $(n-2)$-form
on $M_n$, in which case  the $M_2$-scalar $A$ is shifted by an
arbitrary Minkowski 1-form that is constant on $M_2$.}
 \be
 A_{i_1\dots i_{n-2}} \to
A_{i_1\dots i_{n-2}} + \partial_{[i_1} a_{i_2\dots i_{n-2}]}
\ee
for  a parameter $a$ that is an $(n-3)$-form on $M_n$.  The `pre-field-strength' 2-form
\be\label{GGG}
G_{i_1\dots i_{n-2}} = dA_{i_1\dots i_{n-2}} + \frac{(n-1)}{2}
s^j\partial_{[j}A_{i_1\dots i_{n-2}]}
\ee
is SDiff$_n$ covariant and satisfies the `pre-Bianchi' identity ${\cal D} G \equiv 0$.
However, it is not pre-gauge invariant since
\be
G_{i_1\dots i_{n-2}} \to G_{i_1\dots i_{n-2}} + d\left[
\partial_{[i_1} a_{i_2\dots i_{n-2}]}\right]\, .
\ee
The pre-gauge-invariant and SDiff covariant 2-form is the $M_n$-vector $F^i$, since
\be
eF^i= \varepsilon^{ijk_1\dots k_{n-2}} \partial_j G_{k_1\dots k_{n-2}}\, . \ee We remark that
the expression (\ref{GGG}) is equivalent to \be G_{i_1\dots i_{n-2}}= d A_{i_1\dots
i_{n-2}} - \frac{1}{2\left(n-2\right)!}\, \epsilon_{jk i_1\dots i_{n-2}} \ s^j \wedge
s^k\, ,
\ee
where $\epsilon_{i_1\dots i_n}$ are the components of an $n$-form
$\epsilon$ defined such that
\be\label{defep}
\varepsilon^{i_1\dots i_n} \epsilon_{j_1\dots j_n} = e
\, n! \, \delta^{[i_1}_{j_1} \cdots \delta^{i_n]}_{j_n}\, .
\ee

\subsection{Actions}

Actions that are invariant under local SDiff$_n$ gauge transformations can be constructed from Minkowski space tensors that  are also scalars on $M_n$ via the SDiff covariant derivative. For
example, the local SDiff$_n$ invariant version of (\ref{rigid}) is
\be\label{scalarlag}
{\cal L} =   \oint \! d^n\sigma  \, e\, \left[ -\frac{1}{2}\eta^{\mu\nu} {\cal D}_\mu\phi {\cal D}_\nu\phi
- V(\phi) \right]\, .
\ee

Given at least $n$ scalar fields, potentials may
also be introduced via the Nambu $n$-bracket: a possible SDiff$_n$ invariant potential for  any $n$ scalar fields $(\phi_1,\dots,\phi_n)$ is
\be\label{pot}
{\cal V} =  \oint \! d^n\sigma  \, e \, \left\{\phi_1, \dots ,\phi_n\right\} ^2\, .
\ee

The main obstacle to the construction of  SDiff$_n$ gauge theories
is, for $D>1$,  the difficulty in finding a suitable `kinetic'
term for the  SDiff$_n$ pregauge potential\footnote{For $D=1$ the pregauge potential  is
the Lagrange multiplier for the SDiff$_n$ constraints \cite{Bergshoeff:1988hw}.}. This difficulty
appears insuperable for  $n\ge 4$, so the main case of interest here will be
$n=3$. However, we begin with a review of the $n=2$ case.

\subsubsection{Gauge theories of area-preserving diffeomorphisms}

For $n=2$, the divergence-free constraint on the YM potential $s$
implies, locally on $M_2$, that
\be\label{n=2solve} e
s^i=\varepsilon^{ij}\partial_i A\, , \ee where the scalar $A$ is the
pre-potential 1-form. Using this, we may rewrite the SDiff$_2$
covariant derivative  as \be\label{covdiv2} {\cal D}\phi = d\phi -
\left\{A,\phi\right\}\, , \ee where $\{,\}$ is the Poisson bracket
of functions on $M_2$; i.e. \be\label{bra-f}
 \left\{A,\phi\right\} := e^{-1}\varepsilon^{ij}\partial_i A\, \partial_j\phi\, .
 \ee
We see that the SDiff$_2$ covariant derivative takes the form of a YM covariant derivative if we re-interpret $A$ as a YM potential taking values in the infinite-dimensional Lie algebra of functions on $M_2$ with respect to the Poisson bracket. This algebra is isomorphic to SDiff$_2$ for $M_2=S^2$; for other topologies there is a finite number of divergence-free vector fields that cannot  be written as in (\ref{n=2solve}) but we  ignore these here.

Now consider the Lagrangian density
\be
{\cal L} = - \frac{1}{4} \oint \! d^2\sigma \, e
\eta^{\mu\rho}\eta^{\nu\sigma} G_{\mu\nu}G_{\rho\sigma}\, .
 \ee
where $G$ is the pre-field strength.  Because of the isomorphism
noted above, this is also a YM field-strength for $A$:
\be\label{G=}
G= dA -
\frac{1}{2}\left\{A,A\right\}\, ,
\ee The action is {\it
not} invariant under the pre-gauge transformation $G\to G + da$,
where $a$ is a scalar on $M_2$, but  this just means that the action  includes a Maxwell action 
for a $U(1)$ factor, which may be omitted because it is decoupled from the other fields. 

\subsubsection{Gauge theories of volume-preserving diffeomorphisms}

For $n=3$ the SDiff pre-field-strength 2-form is
\be\label{G3} G_i
=dA_i + s^j\partial_{[j}A_{i]}\, .
\ee
This is not a YM field strength. One might wonder, by analogy with the SDiff$_2$ case,  whether $A_i$ takes values in {\it some} Lie algebra, presumably related to SDiff$_3$, but it appears that such a re-interpretation is not possible \cite{Mkrtchian:1994qc}. We are now dealing with an `exotic' gauge theory.
In view of this, it is not surprising that there is  no longer any way to form a standard YM Lagrangian density.  In any case, $G_i$ is not pre-gauge invariant. However, the Minkowski scalar density
\be
{\cal L} = \oint d^3\sigma \, e\, F^i_{\mu\nu}\, G_i^{\mu\nu}
\ee
is a possible kinetic term; it is both SDiff$_3$ gauge invariant, manifestly, and
pregauge-invariant as a consequence of the constraint (\ref{divfree}).
One may use this term to construct gauge theories that are analogous in many respects to standard Yang-Mills theories; in particular,  one may construct simple supersymmetric gauge theories of volume-preserving diffeomorphisms in dimensions $D=3,4,6,10$.

Here we present the $D=10$ case for which the superpartner to the
gauge prepotential $A_i$ is a Majorana-Weyl spinor that is also a
1-form on $M_3$; the result also applies, {\it mutatis mutandis},
for $D=3,4,6$. Suppressing the Lorentz spinor index,  we denote this
superpartner by  $\chi_i$, and we take $\bar\chi_i$ to be the $D=10$
Majorana conjugate spinor. Let $\Gamma^\mu$ be the $D=10$ Dirac
matrices, and $\Gamma^{\mu\nu}$ the antisymmetrized product of two
of them (with `strength one' convention for antisymmetrization). Now
consider the SYM-like Lagrangian density
 \be\label{cLSYM10} {\cal L}= \oint
d^3\sigma \left[ e\, F^i_{\mu\nu}G_i^{\mu\nu} + i
\varepsilon^{ijk}\partial_i\bar\chi_j\Gamma^\mu\left({\cal
D}_\mu\chi\right)_k \right]\, . \ee Using the pre-Bianchi identity
${\cal D}G \equiv 0$, and the usual $D=10$ Dirac-matrix identities,
one can show that the corresponding action is invariant under the
supersymmetry transformations
\be \label{susySYM10}\delta A_{\mu i} = i
\bar{\epsilon}\Gamma_\mu \chi_i\; , \qquad
  \delta \chi_i= e^{-1}\, \Gamma^{\mu\nu}{\epsilon}\,  G_{\mu\nu \, i}\, ,
\ee
where the parameter $\epsilon$ is a constant anti-commuting Majorana-Weyl spinor.

This construction uses the fact that there is a natural bilinear
inner product  $\langle \; | \;\, \rangle$ on the
space of one-forms on $M_3$: the inner product  of  one-forms
$\omega$ and $\omega^\prime$ is \be \langle \omega | \omega' \rangle
:= \oint \! d^3\sigma\, \varepsilon^{ijk} \omega_i \partial_j
\omega^\prime_k\, . \ee However, this inner product is not positive
semi-definite, and this means that the energy density will not be
positive definite.  A more physical class of  SDiff$_3$ gauge
theories is possible for $D=3$, as we explain in the following
section.

\subsubsection{$n\ge4$}

For $n=4$ the pre-field-strength $G_{ij}$ is an abelian 2-form
potential on $M_4$, and the field-strength $F^i$ is (as always) a
vector. As for $n=3$, there is no way to construct an SDiff$_4$
invariant from products of $F^i$ alone, so $G_{ij}$ must be used too
but the  possibilities are then severely restricted by the
requirement of pre-gauge invariance. In fact, there are no
SDiff$_4$ and pregauge invariants that can be constructed
from $G_{ij}$ and $F^i$ alone, and the same applies for $n>4$.
We will not pursue the possibility that such invariants exist
once additional fields are introduced since we have not found anything
useful in this way.

\section{Conformal SDiff$_3$ gauge theories}
\setcounter{equation}{0}

There is an additional possibility for SDiff$_3$ gauge theories that arises only for $D=3$. Consider first, for  a $D=4$ Minkowski spacetime, the Minkowski  4-form
\be
L_{FG} = \oint d^3\sigma \, e\, F^i \wedge G_i\, .
\ee
This is manifestly SDiff$_3$ gauge invariant, and pregauge invariant as a consequence of the
constraint (\ref{divfree}). One may show that, locally on Minkowski spacetime,
 \be
L_{FG} = d L_{CS}
 \ee
where
\be\label{LCS=SDiff}
L_{CS}  =  \oint \! d^3\sigma\, e\left[ds^i \wedge A_i - \frac{1}{3}
\epsilon_{ijk}s^i \wedge s^j\wedge s^k\right]\, .
\ee
Recall, as a special case of (\ref{defep}), that the alternating tensor $\epsilon_{ijk}$ is defined by
\be
\varepsilon^{ijk} \epsilon_{\ell mn} = 6e\, \delta^{[i}_{\ell}\delta^j_m\delta^{k]}_n\, .
\ee
We note, for future use, that for any variation $\delta A_i$ of $A_i$, one has
\be
\delta L_{CS} = 2 \oint \! d^3\sigma\, e\,  \delta A_i \wedge F^i -
d \left[ \oint \! d^3\sigma\, e s^i \wedge \delta A_i\right] \, .
\ee

\subsection{Chern-Simons-type gauge theories}

We may now use $L_{CS}$ as a Lagrangian 3-form for a $D=3$ Minkowski
spacetime. This yields the Lagrangian density \be {\cal L}_{CS} =
\oint d^3\sigma \, e \epsilon^{\mu\nu\rho} \left[ \left(\partial_\mu
s_\nu^i\right) A_{\rho\, i} - \frac{1}{3} \epsilon_{ijk} s_\mu^i
s_\nu ^j s_\rho^k\right] \, . \ee Omitting a total spacetime
derivative, one has for arbitrary variation $\delta A_{\mu i}$,
\be\label{genvar} \delta L_{CS} = \oint \!d^3\sigma
\left[\varepsilon^{\mu\nu\rho} \delta A_{\mu i} F_{\nu\rho}^i
\right]\, , \qquad F_{\mu\nu}^i := 2\left(\partial_{[\mu} s_{\nu]}^i
+ s_{[\mu}^j\partial_j s_{\nu]}^i\right)\, . \ee One may use this
result to verify that the action is both SDiff$_3$  invariant and,
because of the constraint   (\ref{con-s}),  pre-gauge invariant; it
is also conformal invariant if  the $M_3$ coordinates are inert and
the   pre-potential  1-form $A_i$ is assigned conformal weight zero
(as for the Minkowski-space  exterior derivative $d$). This action
is analogous to the Chern-Simons (CS) term of a $D=3$ YM gauge
theory, but the analogy is not complete because $A_i$ is {\it not} a
YM gauge potential, but rather its pre-potential, and for this
reason we will say that it is of  CS `type'. A peculiarity of this
CS-type term is that it is parity-even rather than parity-odd
because a parity flip in the $D=3$ spacetime can be compensated by a
parity flip of $M_3$.

Suppose that we add to ${\cal L}_{CS}$  the `matter' Lagrangian
density
\be {\cal L}_{mat} = \frac{1}{2}\oint \! d^3\sigma
\, e\, \left({\cal D}\phi\right)^2\, ,
\ee
 In this case, the variation of $A_i$ yields the
SDiff$_3$-invariant equation \be \star  F^i = - J^i \equiv -
\frac{1}{2} e^{-1} \varepsilon^{ijk}
\partial_j \vec\phi  \cdot D \partial_k \vec\phi\, .
\ee
Here we use the language of differential forms in $D=3$ Minkowski space with
$\star$ the Hodge dual operator.

\subsection{SDiff$_3$ $\to$ SDiff$_2$}
\label{subsec:3to2}

Consider the following Lagrangian density
\be\label{DRlag}
{\cal L} = \oint \! d^3\sigma\, e \left[ - \frac{1}{2} \eta^{\mu\nu}\left( {\cal D}_\mu \phi {\cal D}_\nu\phi \right)\right] + \frac{1}{2g}{\cal L}_{CS}\, ,
\ee
where $g$ is an arbitrary non-zero coupling constant.  Let us suppose that the `internal' 3-manifold of this theory takes the form
\be
M_{3} = M_2 \times S^1
\ee
for some 2-manifold $M_2$. In this case  we may split the local $M_3$ coordinates such that
\be
\sigma^i \to \left(\sigma^a, \sigma^*\right), \qquad (a=1,2)
\ee
where $\sigma^a$ are local coordinates for $M_2$, and $\sigma^*$ is a local coordinate for $S^1$, periodically identified with unit period. We also have $e= e_2 e_1$ where $e_2$ is a scalar density on $M_2$, and we may choose  $e_1=1$  without loss of generality, so that $e_2=e$.

If we suppose that $\phi$ is periodically identified then \be \phi
\sim \phi + \sqrt{m} \ee for some mass parameter $m$ since $\phi^2$
has dimensions of mass in fundamental units. The $\phi$ field now
maps the $S^1$ factor of $M_3$ to another
circle, so the $\phi$ field space decomposes into a sum of spaces
with distinct degree for this map. We will focus on the maps of
degree one, for which \be\label{gaugefix} \phi = \sqrt{m}\, \sigma^*
+ \varphi\, , \ee where $\varphi$ is a function on $M_2$ only. The
SDiff$_3$ gauge variation of $\varphi$ is \be \delta_\xi \varphi = -
\xi^a\partial_a \varphi - \sqrt{m} \xi^*\, . \ee This allows us to
partially fix the SDiff$_3$ gauge invariance by choosing \be
\varphi=0\qquad \left( \Rightarrow {\cal D}\phi = \sqrt{m}\,
s^*\right)\, . \ee This restricts us to SDiff$_3$ gauge
transformations with $\xi^*=0$; i.e. the $\xi^a$ transformations,
but these are not yet those of SDiff($M_2$) because $\xi^a$ may
still depend on $\sigma^*$.  This is understandable because all
fields may also still depend on $\sigma^*$.

To proceed, we will now dimensionally reduce by declaring that all fields (other than $\phi$) are independent of $\sigma^*$. This is, of course, equivalent to keeping only the leading term in a
Fourier expansion of all fields. In particular, we have $\partial_* s^*=0$, so the constraint
 (\ref{con-s}) reduces to
\be
\partial_ a \left(e s^a\right)  = 0\, ,
\ee and hence $s^*$ (actually its zero mode on $S^1$) is now
unconstrained. Moreover, the CS-type 3-form  reduces to the sum of an exact 3-form and
the 3-form
 \be
 L_{CS} =  2 \oint \! d^2\sigma \, e s^* \wedge G\, ,
\ee where $G$ is the YM field strength 2-form: \be G\equiv G_* =
dA_* - \frac{1}{2} e^{-1} \varepsilon^{ab} \partial_a A_* \partial_b
A_*  \, . \ee Our starting Lagrangian density (\ref{DRlag}) now
becomes \be {\cal L} =  -\frac{1}{2} \oint \! d^2\sigma \, e\left[ m
\, \eta^{\mu\nu}s_\mu^* s_\nu^*  -  \frac{1}{g}  s_\mu ^*\,
\varepsilon^{\mu\nu\rho} G_{\nu\rho}\right] \, . \ee Eliminating
$s_\mu^*$,  we arrive at the Lagrangian density for an SDiff$_2$
pure YM theory:
\be {\cal L}  = - \frac{1}{4mg^2} \oint \! d^2\sigma
\, e\, G_{\mu\nu}G^{\mu\nu} \, . \ee
As pointed out below (\ref{G=}), this is not pregauge invariant; as derived here, this follows from the fact that the dimensional reduction breaks pre-gauge invariance.

\subsection{${\cal N}=1$ Supersymmetry}

It is straightforward to construct  ${\cal N}=1$ supersymmetric  actions invariant under SDiff$_3$ gauge transformations for $D=3$.  We will need to introduce $2\times 2$ Dirac matrices
$\gamma^\mu$, which we may choose such that
\be
\gamma^{\mu\nu} = \varepsilon^{\mu\nu\rho}\gamma_\rho\, .
\ee
We will also need to introduce the $D=3$ charge conjugation matrix $C$, which is real antisymmetric, and equal to $\gamma^0$ in a real representation for the Dirac matrices. Note that the matrices $C\gamma^\mu$ are symmetric. For a Majorana spinor,  $\lambda$ say, the Dirac conjugate equals the Majorana conjugate, so
\be
\bar\lambda = \lambda^t C\, ,
\ee
where the superfix $t$ indicates `transpose'.

Let us consider first the supersymmetric extension of the `CS' term.
This is \be\label{susyCS} {\cal L}_{CS}^{{\cal N}=1} = {\cal L}_{CS}
-\frac{i}{2} \oint \! d^3\sigma\, \varepsilon^{ijk}\bar\chi_i \,
\partial_j\chi_k\, , \ee where  $\chi_i=dx^\mu \chi_{\mu i}$ is
a Grassmann-odd 1-form on $M_3$ that is also a $D=3$
Minkowski space Majorana spinor (we suppress spinor indices). The
corresponding action is invariant under the infinitesimal
supersymmetry transformations \be\label{vecsusy} \delta A_{\mu \, i}
= \frac{i}{\sqrt{2}} \bar\epsilon \gamma_\mu \chi_i\, , \qquad
\delta\chi_i = -\frac{1}{\sqrt{2}} \gamma^{\mu\nu} \epsilon\,
G_{\mu\nu\, i}\, , \ee where $G_{\mu\nu\, i}$ are the components of
the pre-field-strength 2-form $G_i$, and $\epsilon$ is a constant
anticommuting Majorana spinor parameter. The coefficient
$1/\sqrt{2}$ is introduced here for later convenience.

We may couple to this CS-type theory any number of scalar multiplets with component fields that are   scalars on $M_3$. For simplicity, we consider a single scalar multiplet with scalar field $\phi$ and two-component Majorana spinor field $\psi$.  Consider the Lagrangian density
\be
{\cal L}_0 = -\frac{1}{2} \oint \! d^3\sigma \, e\left[ \eta^{\mu\nu}{\cal D}_\mu\phi {\cal D}_\nu\phi + i\bar\psi \gamma^\mu {\cal D}_\mu \psi  + \left(W'\right)^2 - i W'{}' \bar\psi \psi \right]\, ,
\ee
for any real (superpotential) function $W(\phi)$.  This is not supersymmetric by itself, but the Lagrangian density
\be\label{matter}
{\cal L}_{matter}^{{\cal N}=1} = {\cal L}_0  -\frac{i}{\sqrt{2}}  \oint \! d^3\sigma\, \varepsilon^{ijk} \left(\bar\psi \, \partial_j\chi_k\right) \partial_i\phi \, ,
\ee
 is invariant under the combined transformations of (\ref{vecsusy}) and
\be
\delta \phi = i\bar\epsilon \psi\, , \qquad
\delta\psi = \left(\gamma^\mu {\cal D}_\mu \phi + W' \right)\epsilon\, .
\ee

If we now add these two ${\cal N}=1$ supersymmetric Lagrangian
densities, introducing a coupling constant $g$ to allow for
different relative weights,  we have
 \be
 {\cal L}= {\cal L}_0 + \frac{1}{2g}{\cal L}_{CS}
- i \oint \! d^3\sigma \varepsilon^{ijk} \left[ \frac{1}{4g}
\bar\chi_i \partial_j\chi_k + \frac{1}{\sqrt{2}} \bar\chi_i
\partial_j \left(\psi\partial_k\phi\right)\right] \, .
 \ee
The $\chi_i$ equation of motion  determines $\chi_i$ only up to a
total $M_3$ derivative because this is clearly a gauge invariance of
the action; we may fix this gauge such that
 \be
\chi_i = -\sqrt{2} g \psi\,  \partial_i\phi \, .
\ee
The net result is the Lagrangian density
\be
{\cal L} = -\frac{1}{2} \oint \! d^3\sigma\,
e\left[\eta^{\mu\nu}{\cal D}_\mu\phi {\cal D}_\nu\phi +i \bar\psi
\gamma^\mu {\cal D}_\mu\psi  + \left(W'\right)^2 - i W'{}' \bar\psi
\psi \right] + \frac{1}{2g}{\cal L}_{CS}\, .
\ee
This is invariant,
omitting a total  spacetime derivatives, under the infinitesimal
supersymmetry transformations
\be \delta \phi = i\bar\epsilon \psi,
, \qquad \delta\psi  = \left(\gamma^\mu {\cal D}_\mu\phi  + W'
\right)\epsilon\, , \qquad \delta A_{\mu i} = -ig \left(\bar\epsilon
\gamma_\mu\psi \right) \partial_i \phi\, .
\ee
The `CS' term is
essential for the invariance, and also needed is the Fierz identity
\be
\gamma^\mu d\chi \left(d\bar\chi \gamma_\mu d\chi\right)  \equiv 0\, .
 \ee

\subsection{Superspace}\label{subsec:superspace}

We now aim to recover the above model using superspace techniques.
We begin by writing the superspace exterior derivative as \be d=
E^\alpha D_\alpha + E^\mu\partial_\mu \ee where $(E^\alpha,E^\mu)$
are a basis of 1-forms on superspace such that the 2-component
Majorana spinor derivative $D_\alpha$ has the anti-commutator
 \be
 \left[
D_\alpha, D_\beta\right]_+ = 2i
\left(C\gamma^\mu\right)_{\alpha\beta}
\partial_\mu\, .
 \ee
The fields $(\phi,\chi)$ combine to form a single {\it superfield}
$\phi$ such that $\sqrt{2} D_\alpha\phi = \chi_\alpha$. As is
customary, we use the same symbol to denote both a superfield and
its first component since when these components are defined in terms
of spinor derivatives (rather than by  superfield expansion) each
component equation may be interpreted as a superfield equation.

SDiff gauge fields are introduced via the SDiff covariant superspace
exterior derivative
\be
{\cal D} = d + {\cal L}_\Sigma \, ,  \qquad
\partial_i \left(e\Sigma^i\right)=0\, ,
\ee where \be\label{Sigma} \Sigma^i  = E^\alpha\varsigma_\alpha^i +
E^\mu  s_\mu^i  \, ,
 \ee
so that
 \be\label{cptsD}
 {\cal D}=E^\alpha {\bb
D}_\alpha + E^\mu {\cal D}_\mu\, ,
\ee
where, for example,
\be {\bb D}_\alpha \phi = D_\alpha \phi  + \varsigma_\alpha{}^i
\partial_i \phi\, , \qquad {\cal D}_\mu\phi =
\partial_\mu\phi + s_\mu{}^i \partial_i \phi\, .
\ee The
components $(\varsigma,s)$ of the superspace SDiff potential
$\Sigma$, both of which are superfields, are related by the
requirement that
 \be
  \left[{\bb D}_\alpha, {\bb D}_\beta\right]_+ =
2i \left(C\gamma^\mu\right)_{\alpha\beta} {\cal D}_\mu\, ,
\ee
which implies that
 \be
 \bb{D}_{(\alpha}\varsigma_{\beta)}^i ={D}_{(\alpha}\varsigma_{\beta)}^i + \varsigma_{(\alpha}^j \partial_j\varsigma_{\beta)}^i
= i \left(C\gamma^\mu\right)_{\alpha\beta} s_\mu^i\, .
\ee
Using this equation, one may show that the  `matter' Lagrangian density of
(\ref{matter}) is reproduced, on elimination of auxiliary fields, by
the {\it superspace}  Lagrangian density
 \be\label{LsuperM}
{\cal L}_{matter} = -\frac{1}{2} \oint\! d^3\sigma\,e \left[ \bar {\bb
D}\phi\, {\bb D}\phi - 4i {W}(\phi)\right]\, .
\ee
To verify this, one must use the superspace integration measure $d^3x
{1\over 8}[\bar{\bb D}, {\bb D}]$.

To write the superspace Lagrangian for the CS-type term we first
solve the divergence-free constraint on $\Sigma$ by writing
\be
\Sigma^i = e^{-1} \varepsilon^{ijk} \partial_j \Lambda_k \, ,
\ee
where $\Lambda_i$ is the superspace pre-potential; in terms of its
(superfield) components $(\lambda_i,A_i)$, we have \be \varsigma^i =
e^{-1} \varepsilon^{ijk} \partial_j \lambda_k \, , \qquad s^i =
e^{-1} \varepsilon^{ijk} \partial_j A_k\, ,
 \ee
where
 \be \label{345}D_{(\alpha}\lambda_{\beta)\, i} =
i\left(C\gamma^\mu\right)_{\alpha\beta} A_{\mu\, i}\, . \ee

Next, we introduce the {\it superspace} SDiff$_3$ field-strength 2-form
\be F^i = d\Sigma^i + \Sigma^j\partial_j \Sigma^i\, . \ee This can
be written, locally on $M_3$, in terms of a {\it superspace}
pre-field-strength $G_i$ as
 \be F^i = e^{-1}\varepsilon^{ijk}
\partial_j G_k \, , \qquad G_i = d\Lambda_i +
\Sigma^j\partial_{[j}\Lambda_{i]}\, .
\ee
One may now show that (\ref{345}) is equivalent to
\be
G_{i\; \alpha\beta}=0\, ,
\ee
which is the pre--field strength analog of the standard Yang--Mills superspace constraints.
This follows from
 \begin{eqnarray}
 \Lambda_i  &=& E^\mu A_{\mu i} + E^\alpha \lambda_{\alpha i}\, , \qquad
 D =  E^\mu D_{\mu} + E^\alpha D_{\alpha}\, , \qquad
 \Sigma_i =  E^\mu A_{\mu i} + E^\alpha \lambda_{\alpha i} \nonumber \\
 G_i &=& {1\over 2} E^\alpha\wedge  E^\beta
G_{\alpha\beta i}+  E^\alpha\wedge  E^\mu G_{\mu \alpha i}+
\frac{1}{2}E^\mu\wedge  E^\nu G_{\nu\mu i}\, ,
\end{eqnarray}
after  taking into account that
\be
dE^a= -2iE^\alpha\wedge E^\beta (C\gamma^\mu)_{\alpha\beta}\, .
\ee

The {\it superspace} 4-form $\oint\! d^3\sigma \, e F^iG_i$ is
both SDiff$_3$ and pregauge invariant, but we cannot use it to construct
directly the superspace integrand for the CS-type term. However,
using the techniques of \cite{Gates:1983nr,Zupnik:1988en} we may map
the `CS' superspace 3-form to the CS-type Lagrangian density
 \be
 {\cal L}  = -4i \oint\! d^3\sigma e\left[ \bar{W}^i
\lambda_i - {1\over 12} \epsilon_{ijk} \bar\varsigma^i
\gamma^\mu\varsigma^j s_\mu{}^k \right]\, ,
\ee
where\footnote{This
quantity arises as the spinor field strength $F_{\alpha
\mu}{}^i= i(\gamma_\mu W^i)_\alpha$; it is the SDiff$_3$
counterpart of the spinorial SYM field strength, and it has its own
pre-field strength ${\cal W}_i$, defined by
$W^i={e}^{-1}\varepsilon^{ijk}\partial_j{\cal W}_k$.}
\be
W^i=-{i\over 2} \gamma^\mu {\cal D}_\mu \varsigma^i + {i\over 4}
\bar{\bb{D}}\bb{D}  \varsigma^i \, . \ee One may verify that this
reproduces  (\ref{susyCS}) in the Wess-Zumino gauge.

\section{BLG}
\setcounter{equation}{0}

Let $\phi^I$ ($I=1,\dots 8$) be a $Spin(8)$ ${\bf 8}_v$-plet of real scalar fields, and $\psi_A$ ($A=1,\dots 8$) a $Spin(8)$ ${\bf 8}_s$-plet of Majorana anticommuting  $Sl(2;\bR)$ spinor fields, both on the cartesian product of 3-dimensional Minkowski spacetime with some  $3$-dimensional closed manifold without boundary, $M_3$. Let $\rho^I$ be the $8\times 8$ $Spin(8)$ `sigma' matrices, and $\tilde\rho^I$ their transposes, as in \cite{Bandos:2008fr}. Note that
\be
\rho^{IJ} := \rho^{[I}\tilde\rho^{J]}
\ee
is {\it antisymmetric} in its spinor indices. We also define
\be
\tilde \rho^{IJK} := \tilde\rho^{[I} \rho^{JK]}\, ,
\qquad \rho^{IJKL} := \rho^{[I}\tilde\rho^{JKL]}\, .
\ee

Now consider the following Lagrangian density
\begin{eqnarray}\label{Lagzero}
{\cal L}_{M2} &=& \oint \!d^3\sigma\, \left[ -\frac{1}{2}e \left| {\cal D}\phi\right|^2 - \frac{i}{2} e\, \bar\psi \gamma^\mu {\cal D}_\mu \psi +  \frac{ig}{4} \varepsilon^{ijk} \partial_i\phi^I\partial_j\phi^J \left(\partial_k\bar\psi \rho^{IJ}\psi\right) \right. \nn
&&\qquad  \qquad \left.
- \frac{g^2}{12} e \left\{\phi^I,\phi^J,\phi^K\right\}^2 \right]
+ \ \frac{1}{2g}{\cal L}_{CS}\, ,
\end{eqnarray}
where $g$ is a real dimensionless parameter, and $Spin(8)$ indices
are suppressed. This Lagrangian density varies into a total
spacetime derivative under the following infinitesimal supersymmetry
transformations with ${\bf 8}_c$-plet constant anticommuting spinor
parameter $\epsilon_{\dot A}$ ($\dot A=1,\dots,8$):
\begin{eqnarray}\label{trans}
\delta\phi^I &=& i \bar\epsilon \tilde\rho^I \psi\, , \qquad
\delta A_{\mu i} = -ig \left(\bar\epsilon\gamma_\mu \tilde\rho^I\psi\right) \partial_i\phi^I\, , \nn
\delta\psi  &=& \left[\gamma^\mu  \rho^I {\cal D}_\mu \phi^I -
 \frac{g}{6} \left\{\phi^I,\phi^J,\phi^K\right\} \rho^{IJK}\right] \epsilon\, .
\end{eqnarray}
To verify this, one needs the `fundamental' identity,  and the Fierz identity
\be\label{Fierz}
\tilde\rho^J \gamma^\mu d\psi \left(d\bar\psi \gamma_\mu d\psi\right) -
\tilde\rho^I d\psi \left(d\bar\psi \rho^{IJ} d\psi\right) \equiv 0 \, .
\ee
If all the fields of this model are expanded in harmonics on
$M_3$  then $L$ becomes the sum of a  Lagrangian $L_0$
describing the centre of mass motion of the M2 condensate and a
remainder that describes the `internal' dynamics. The centre-of-mass
fields come from the constant harmonic on $M_3$. There
is no contribution of the constant harmonic to $s^i$ since this is a
vector on $M_3$ (see e.g. \cite{Dowker:1990iy}), so the centre of mass fields are those of
a single ${\cal N}=8$ supermultiplet, with no interactions.

\subsection{Fierz identity}

Let us pause to prove (\ref{Fierz}). The  LHS can be rewritten by a Fierz rearrangement as
\be\label{LHS}
LHS = \frac{1}{16} d\bar\psi {\cal O}^A d\psi \left[ \tilde\rho^J\gamma^\mu {\cal O}^A \gamma_\mu -
\tilde \rho^I {\cal O}^A \rho^{IJ}\right] d\psi\, ,
\ee
where the overall sign  is plus because $d\psi$ is commuting,  and
${\cal O}^A$ is a complete set of the $16 \times 16$ matrices  formed by tensor products of
$(1,\gamma^\mu)$ with $(1,\rho^{IJ},\rho^{IJKL})$. Actually, the only matrices of this type which contribute are those for which $C{\cal O}^A$ is symmetric (because $d\psi$ is commuting). This means that we have only to consider
\be
\gamma^\mu \otimes 1 \, , \qquad 1 \otimes \rho^{IJ}\, , \qquad \gamma^\mu \otimes \rho^{IJKL}\, .
\ee
It should be clear that the first two of these will produce terms of a type that already appear on the LHS of  (\ref{Fierz}) whereas the third does not. However, this `third'  matrix gives a contribution proportional to
\be
d\bar\psi \gamma^\nu \rho^{KLMN}d\psi\left[
- \tilde\rho^J \rho^{KLMN}  - \tilde\rho^I \rho^{KLMN}\rho^{IJ} \right]  \gamma_\nu d\psi
\ee
where we have used $\gamma^\mu\gamma_\nu\gamma_\mu \equiv -\gamma_\nu$.  But this contribution is zero as a consequence of the identities
\be
\rho^{IJ} \equiv \rho^I \tilde \rho^J - \delta^{IJ}\, , \qquad \tilde\rho^I \rho^{KLMN} \rho^I \equiv 0\, .
\ee
This cancelation means that  we now have
\be\label{remain}
LHS = \frac{1}{16} d\bar\psi \gamma^\nu d\psi \left[ - \tilde\rho^J -
 \tilde\rho^I\rho^{IJ}\right] \gamma_\nu d\psi -
  \frac{1}{32} d\bar\psi \rho^{KL} d\psi \left[ 3 \tilde\rho^J\rho^{KL} -
  \tilde\rho^I\rho^{KL}\rho^{IJ}\right] d\psi\, .
\ee
The overall minus sign of the second term arises because matrices like $\rho^{12}$ square to {\it minus} the identity, and the additional factor of $1/2$ compensates for the double counting implied by the index summation convention.  Using the identities
\be
\tilde\rho^I\rho^{IJ} \equiv 7 \tilde\rho^J\, , \qquad \tilde\rho^I \rho^{KL} \rho^{IJ} \equiv
4\rho^{KL}\tilde \rho^J - \tilde\rho^J\rho^{KL}\, ,  \qquad [\tilde\rho^J,\tilde\rho^{KL}] = 4 \delta^{J[K}\tilde\rho^{L]}\, ,
\ee
we now find that
\be
LHS = -\frac{1}{2} \left(d\bar\psi \gamma^\nu d\psi \right) \tilde \rho^J \gamma_\nu d\psi
+ \frac{1}{2} \left(d\bar\psi \rho^{IJ} d\psi\right) \tilde\rho^I d\psi \equiv -\frac{1}{2} LHS\, ,
\ee
from which it follows that  $LHS=0$, which is just the Fierz identity (\ref{Fierz}).

Another way to prove the Fierz identity is to show that it follows from the $D=11$ Dirac-matrix
identity  that allows the construction of the $D=11$ supermembrane  \cite{Bergshoeff:1987cm}.  To see this, first  write this $D=11$  identity in the form
\be
 \Gamma^{MN} d\Psi   \left(d\bar \Psi \Gamma_M d\Psi\right)
+  \Gamma_M d\Psi \left(d\bar\Psi \Gamma^{MN} d\Psi \right) \equiv
0\, , \ee
where $\Psi$ is an anticommuting D=11 Majorana spinors, and
$\Gamma^M$ are the $D=11$ Dirac  matrices. Next,  split the 11-vector index $M\to (\mu,I)$,
breaking $Spin(1,10)$ to $Sl(2;\bR)\times Spin(8)$,  and consider the $I$
component of the D=11 identity for \be\label{PsiMW=} \Psi =
\pmatrix{\psi \cr 0} \, , \ee where the 16-component $\psi$
transforms as the real $({\bf 2},{\bf 8}_c)$ of $Sl(2;\bR)\times
Spin(8)$. This yields the identity (\ref{Fierz}).

\subsection{Superconformal invariance}

The Noether current corresponding to the invariance of ${\cal L}_{M2}$ under the supersymmetry transformations (\ref{trans}) is
\be
{\cal S}_{Noether}^\mu = \oint \!d^3\sigma \left[ \gamma^\nu \gamma^\mu \tilde\rho^I \psi {\cal D}_\nu\phi^I - \frac{g}{6} \gamma^\mu \tilde\rho^{IJK}\psi \left\{\phi^I,\phi^J,\phi^K\right\} \right]
\ee
but we may add to this any vector spinor that is {\it identically} divergence-free. Consider, in particular, the `improved'  supersymmetry current
\be
{\cal S}^\mu = \oint \!d^3\sigma \left[ \gamma^\nu \gamma^\mu \tilde\rho^I \psi {\cal D}_\nu\phi^I - \frac{g}{6} \gamma^\mu \tilde\rho^{IJK}\psi \left\{\phi^I,\phi^J,\phi^K\right\}
-\frac{1}{2} \tilde\rho^I \gamma^{\mu\nu} \partial_\nu \left(\phi^I\psi\right)\right]\, ,
\ee
which differs from the Noether current by the addition of the final term, which is identically divergence-free. As a consequence of this addition,  one finds that  the $\psi$ equation of motion implies that
\be
\gamma_\mu {\cal S}^\mu =0\, .
\ee
This implies that ${\cal S}^\mu$ is part of a supermultiplet that contains the `improved', because trace-free, energy-momentum stress tensor, which in turn implies that the model  is superconformal invariant.

Note that $g$  cannot be set to zero in the action because of the CS term. In fact, $|g|$ may be set to unity without loss of generality because, when $|g|\ne1$, the scaling
\be
A\to |g|^{2/3}A\, , \qquad \sigma^i \to |g|^{1/3}\sigma
\ee
has the effect of  taking $|g|\to 1$, except for an overall factor coming from  the $\oint \!d^3\sigma$ integral. The choice of sign of $g$ is presumably related to whether we wish to describe a condensate of M2-branes or anti-M2-branes.

\subsection{Equations of motion and the free-field limit}

The equations of motion are
\begin{eqnarray}\label{EofM}
0 &=& {\cal D}^\mu {\cal D}_\mu \phi^I - i \frac{g}{2} e^{-1}\,
\varepsilon^{ijk}\partial_i\phi^J \partial_j\bar\psi \rho^{IJ}
\partial_k \psi +
\frac{g^2}{2}\left\{\left\{\phi^I,\phi^J,\phi^K\right\},
\phi^J,\phi^K\right\}\, ,  \nonumber \\
0 &=& \gamma^\mu
{\cal D}_\mu \psi + \frac{g}{2} \rho^{IJ}
\left\{\phi^I,\phi^J,\psi\right\}\, ,  \\
0 &=&
{1\over 2}\varepsilon^{\mu\nu\rho}  F_{\nu\rho}^i  + g \, e^{-1}\,
\varepsilon^{ijk}\left[ \partial_j\phi^I {\cal D}^\mu
\partial_k \phi^I -\frac{i}{2} \partial_j\psi\gamma^\mu
\partial_k\psi\right]\, . \nonumber
\end{eqnarray}
Although we were unable to set $g=0$ in the action, this can be done in the equations of motion.
The result is that $F=0$, so that $s$  is pure gauge. We may then choose a gauge for which $s=0$, at which point we see that we have free field equations for $\phi$ and $\psi$. These equations are those of a supersymmetric theory with transformations given by (\ref{trans}) for $g=0$ and $s=0$.

\subsection{M2  boundaries}

Bosonic configurations that preserve susy have a spinor $\epsilon$ that obeys
 \be\label{susypres}
 \rho^I \gamma^\mu {\cal D}_\mu\phi^I
\epsilon = \frac{g}{6}\left\{\phi^J,\phi^K,\phi^L\right\}
\rho^{JKL}\epsilon\, .
 \ee
Let us choose $M_3=S^3$ and consider bosonic configurations for
which \be \phi^a = f(x^1)X_a(\sigma)\, , \qquad a=1,2,3,4 \ee where
$X_a$ are the functions that map $M_3$ to the unit  3-sphere, as discussed in the
action for  general $n$: i.e.
\be
\sum_{a=1}^4 X_a^2 =1\, , \qquad
\left\{X_a, X_b,X_c\right\} = \epsilon^{abcd} X_d\, .
\ee
The field equation for the gauge potential $A$ is then solved by
$s=0$, and the $\phi$ equation reduces to \be f'{}'   = 3g^2f^5\, .
\ee This is solved by solutions of \be f' = -gf^3\, , \ee which
preserve 1/2 supersymmetry since the supersymmetry preservation
condition (\ref{susypres}) for such solutions reduces to \be
f\left(1- \gamma^1\rho_\star\right)\epsilon=0 \ee where the matrix
$\rho_\star$, defined by
\be \frac{1}{6}\epsilon^{abcd}\rho^{bcd} =
\rho^a \rho_\star \, , \ee squares to the identity.  Thus, we have 1/2
supersymmetric solutions\footnote{Generic supersymmetric
configurations have been classified in \cite{Jeon:2008bx,
Jeon:2008zj}.} of the form \be \phi^a =
\frac{X_a(\sigma)}{\sqrt{2gx^1}} \ee with all other fields equal to
zero \cite{Bagger:2007jr}.

Let $T$ be the M2 tension, and define the rescaled field with
dimensions of length, \be \Phi^a =\phi^a /\sqrt{T}\, . \ee Because
$\sum_a (X_a)^2=1$, we have \be \sum_{a=1}^4 \left(\Phi^a\right)^2
= \left(1/\sqrt{2gT x^1}\right)^2\, , \ee which shows that at fixed
$x^1$ we have a 3-sphere of radius $r =1/\sqrt{2gT x^1}$. This goes
to infinity as $x^1\to 0$, which means that the M2-branes have
expanded to a planar 5-brane at $x^1=0$. From the 5-brane
perspective, there is a membrane `spike'   with 3-sphere cross
section such that
\be
x^1 = \frac{1}{2gT\, r^2} \, .
\ee
This solves the
Laplace equation on $\bE^4$, in polar coordinates
$(r,\theta,\varphi,\xi)$. In other words we have a solution analogous to that found in
\cite{Howe:1997ue} representing M2-branes ending on an M5-brane. The
5-brane tension was computed in \cite{Bagger:2007vi} and shown to equal the
M5-brane tension.

\subsection{D2 condensate from M2 condensate}

Recalling that a  D2-brane of IIA superstring theory is just an
M2-brane of M-theory compactified on a circle
\cite{Townsend:1995af}, we should expect some analogous relation
between the D2 and M2 condensates. The former is an $N\to\infty$
limit of a maximally supersymmetric $D=3$ YM gauge theory with gauge
group $SU(N)$; as explained in the introduction, this limit yields
an SDiff$_2$  YM theory, so a  D2-condensate is described (at low energy) by an ${\cal N}=8$
supersymmetric $D=3$ YM gauge theory with  gauge group SDiff$_2$. We
shall now exploit our earlier discussion of subsection
\ref{subsec:3to2} to show how this theory is obtained from the BLG
SDiff$_3$  gauge theory.

As in subsection \ref{subsec:3to2}, we choose $M_3=M_2\times S^1$,
such that $\sigma^a$ are local coordinates for $M_2$ and $\sigma^*$
is an coordinate for the $S^1$ factor, periodically identified with
unit period, and we take the density $e$ to be a volume density for
$M_2$. We then set \be \phi^I=(\phi^{\cal I},\phi^8) \qquad  ({\cal
I}=1,\dots,7) \ee and periodically identify $\phi^8$ with period
$\sqrt{m}$. Again following subsection \ref{subsec:3to2}, we
partially fix the SDiff gauge invariance by choosing \be\label{gf8}
\phi^8 = \sqrt{m}\,  \sigma^*\, , \ee and we then choose to consider
only the zero modes on $S^1$ of all other fields.  Let us apply this
generalized dimensional reduction\footnote{It is actually a
supersymmetry-preserving variant of Scherk-Schwarz reduction similar
to that considered in \cite{Bergshoeff:1996ui}.} to the BLG theory.
Relative to the discussion of  subsection \ref{subsec:3to2}, there
are several new ingredients. Firstly, there are an additional 7
scalar fields, for which \be {\cal D}\phi^{\cal I} \to D\phi^{\cal
I}  := d\phi^{\cal I} - \left\{A,\phi^{\cal I}\right\}\, , \qquad (A
:= A_*) \ee which is the YM covariant derivative for the group
SDiff$_2$, realized via the Poisson bracket $\{,\}$ of functions on
$M_2$, as defined in (\ref{bra-f}).  The SDiff$_3$
covariant derivative of the spinor field $\psi$ similarly reduces to
an SDiff$_2$ YM derivative. Secondly, there is a scalar potential
 \be V := \frac{g^2}{12}
\left\{\phi^I,\phi^J,\phi^K\right\}^2 \to \frac{mg^2}{4}
\left\{\phi^{\cal I},\phi^{\cal J}\right\}^2\, .
 \ee
Finally there is the Yukawa-type term \be i\frac {g}{4}
\varepsilon^{ijk}
\partial_i\phi^I\partial_j\phi^J \left(\partial_k\bar\psi \rho^{IJ}
\psi\right) \to i\frac{g\sqrt{m}}{2} \varepsilon^{ab}
\partial_a \phi^{\cal I} \left(\partial_b\bar\psi  \rho^8 \tilde\rho^{\cal I} \psi\right) \, .
\ee Here we have split the eight $SO(8)$ sigma-matrices into
$\rho^8$ and the seven $SO(7)$  sigma matrices  $\rho^{\cal I}$.
We thus find that
\begin{eqnarray}\label{D2con}
{\cal L}_{M2}  \to {\cal L}_{D2} &:=& \oint \! d^2\sigma\, e\left[ -\frac{1}{2}D^\mu\phi^{\cal I} D_\mu\phi^{\cal I}  -\frac{1}{4mg^2} G_{\mu\nu} G^{\mu\nu}  - \frac{mg^2}{4} \left\{\phi^{\cal I},\phi^{\cal J}\right\}^2\right. \nonumber \\
&& \left. -  \frac{i}{2}\bar\psi \gamma^\mu D_\mu \psi
+ i\frac{g\sqrt{m}}{2} \varepsilon^{ab}
\partial_a \phi^{\cal I} \left(\partial_b\bar\psi  \rho^8 \tilde\rho^{\cal I} \psi\right) \right]\, .
\end{eqnarray}
The corresponding  action  is invariant under transformations of
${\cal N}=8$ supersymmetry that may be deduced\footnote{In
principle,  it  is necessary to include a compensating $S^1$-diffeomorphism to maintain
the partial gauge choice (\ref{gf8}), but this has no effect on the fields appearing in
(\ref{D2con}) as these are $\sigma^*$-independent $M_2$-scalars.}
 from (\ref{trans}).  As the SDiff$_2$ gauge group may be viewed as an $N\to\infty$ limit
 of $SU(N)$, it is natural to interpret  ${\cal L}_{D2}$ as the Lagrangian
 density describing the low-energy dynamics of a D2-condensate, related to the
 M2-condensate by reduction on the M-theory circle.

As a further check, we will now show that ${\cal L}_{D2}$ is the
dimensional reduction on $T^7$  of a  $D=10$ SYM theory with
SDiff$_2$ gauge group. The fields of the latter theory are a
Minkowski 1-form  potential $A_m$ ($m=0,1,\dots,9$) and a
Majorana-Weyl spinor  $\Psi$, both scalars on $M_2$.  Let $\Gamma^m$
be $D=10$ Dirac matrices and $\bar\Psi$ the $D=10$
Majorana-conjugate of $\Psi$. The $D=10$ Lagrangian  density is
 \be\label{cL10SD2} {\cal L}_{10} = \frac{1}{g_{10}^2} \oint\!
d^2\sigma\, e \left[ -\frac{1}{4} G_{mn} G^{mn}
 -  \frac{i}{2}\bar{\Psi} \Gamma^m D_m \Psi \right]\, ,
 \ee
where $g_{10}$ is a 10-dimensional coupling constant, and \be G_{mn}
= 2\partial_{[m} A_{n]}  - \left\{ A_m,A_n \right\}\, , \qquad D_m
\Psi = \partial_m\Psi  - \left\{A_m, \Psi\right\} \, . \ee In
fundamental units, the mass dimensions are
\be \left[ A
\right] = 1\, \qquad \left[\Psi \right] =\frac{3}{2}\, , \qquad
\left[ g_{10} \right] = -3\, . \ee It may be verified that ${\cal
L}_{10}$ varies into a total spacetime derivative under the
following infinitesimal supersymmetry transformations
  \be
  \delta A_m =  i  \bar \epsilon \Gamma_m \Psi\, , \qquad \delta
\Psi = -\frac{1}{2}\,  \Gamma^{mn}\epsilon \, G_{mn}\, .
 \ee
To dimensionally reduce to $D=3$, we choose  real $D=10$ Dirac
matrices of the form \be \Gamma^\mu = \gamma^\mu\otimes \gamma^8\, ,
\qquad \Gamma^{\cal I} = \bI_2 \otimes \gamma^{\cal I}\, , \ee where
$(\gamma^{\cal I},\gamma^8) = \gamma^I$ are the $16\times 16$
$SO(8)$ Dirac matrices, which we may write as \be \gamma^I =
\pmatrix{0 & \rho^I\cr \tilde\rho^I &0}\, . \ee
In this basis, the Majorana--Weyl spinor $\Psi$ takes the form of
(\ref{PsiMW=}). Dimensional reduction to
$D=3$ of the Lagrangian density ${\cal L}_{10}$ now yields ${\cal
L}_{D2}$ if we set
\be \frac{{\rm Vol}
(T^7)}{g_{10}^2} = \frac{1}{mg^2}
\ee
and
\be A_{\cal I} = \sqrt{m}\, g\, \phi^{\cal I}\, ,
\qquad \Psi = \sqrt{m}\, g\, \psi \, . \ee
Note that this implies that $\left[ \phi\right]= 1/2$ and
$\left[\psi\right] =1$,  as expected for $D=3$ fields.

Naturally, if we compactify from $D=10$ on $T^6$, rather than $T^7$, we get a $D=4$ ${\cal N}=4$
SDiff$_2$ gauge theory, and  $S^1$-compactification of this theory  yields the $D=3$ ${\cal N}=8$
SDiff$_2$ gauge theory.

\subsection{${\cal N}=8$ superfields}
\label{subsec:on-shell}

Following the original version of this paper, an ${\cal N}=8$ superfield formulation of the Nambu bracket BLG field equations was found \cite{Bandos:2008df}; it  consists of two coupled ${\cal N}=8$ superfield  equations for the SDiff gauge field and the scalar superfield that is also a scalar on the three-dimensional manifold $M_3$. We shall now review this formulation.

We may define an SDiff$_3$-covariant exterior derivative ${\cal D}$
on ${\cal N}=8$ superspace exactly as for ${\cal N}=1$ superspace,
by introducing  the $M_3$-vector-valued 1-form potential
$\Sigma^i$, which is now an  ${\cal N}=8$ superfield: we now have
the following decomposition generalizing (\ref{cptsD}):
\be
{\cal D}
= E^{\alpha \dot A} \bb D_{\alpha \dot A} + E^\mu {\cal D}_\mu\, ,
\ee
where
 \be\label{bbDs}
 \bb D_{\alpha\dot A} = D_{\alpha\dot
A} + \varsigma_{\alpha\dot A}{}^i
\partial_i\, , \qquad {\cal D}_\mu = \partial_\mu + s_\mu{}^i
\partial_i\, .
\ee Here $D_{\alpha\dot A}$ is the standard ${\cal N}=8$ superspace
spinor derivative, and $\varsigma_{\alpha \dot A}{}^i$ is
the ${\bf 8}_c$-plet of superpartners to the SDiff$_3$ gauge field
$s_\mu{}^i$; we shall confirm this below by showing that their
respective field strengths are components of a field-strength
superfield.

When acting on an $M_3$-scalar,
\be\label{defF}
{\cal D}^2 = F^i\partial_i\, ,
\ee
where $F^i$ is the $M_3$ vector-valued ${\cal N}=8$  field strength 2-form superfield.
 Equivalently, but in  terms of the components of  ${\cal D}$ and $F^i$, we have
\begin{eqnarray}\label{D21}
{} [\bb{D}_{\alpha \dot{A}} , \bb{D}_{\beta \dot{B}}]_+  &=& 2i \delta_{\dot{A}\dot{B}}
(C\gamma^\mu)_{\alpha\beta}{\cal D}_\mu +
F_{\alpha \dot{A}\; \beta \dot{B}}{}^i\,
\partial_i \\ \label{D21+}
\left[{\bb D}_{\alpha \dot A}, {\cal D}_\mu \right] &=& F_{\alpha \dot A\, \mu}{}^i
\partial_i  \\ \label{D21++}
\left[{\cal D}_\mu, {\cal D}_\nu \right] &=& F_{\mu\nu}{}^i \,
\partial_i\, .
\end{eqnarray}
Following  \cite{Bandos:2008df}  we impose the constraint
\be
F^i_{\alpha \dot{A}\; \beta \dot{B}} = 2iC_{\alpha\beta} \, W_{\dot{A}\dot{B}}{}^i\, ,
\ee
where $W_{\dot{A}\dot{B}}{}^i$ is in the  {\bf 28} of SO(8); it is also divergence-free, so
\be
W_{\dot{A}\dot{B}}{}^i=-W_{\dot{B}\dot{A}}{}^i\, , \qquad
\partial_i (e \,W_{\dot{A}\dot{B}}{}^i)=0\, .
\ee
Using the Jacobi identity
\be\label{Jacobi}
\left[ \bb D_{\alpha \dot A},  \left[ \bb D_{\beta\dot B}, \bb D_{\gamma \dot C}  \right]_+ \right]_-
 + \left[ \bb{D}_{\beta \dot{B}}  ,
 \left[ \bb{D}_{\gamma \dot{C}}  ,  \bb{D}_{\alpha \dot{A}}
 \right]_+\right]_-  + \left[ \bb{D}_{\gamma \dot{C}} ,
 \left[ \bb{D}_{\alpha \dot{A}}  ,  \bb{D}_{\beta \dot{B}} \right]_+ \right]_-
 \equiv0\, ,
 \ee
 one finds that
 \be
 F_{\alpha \dot A\, \mu}{}^i  = i  \left(\gamma_\mu W_{ \dot A}{}^i \right)_\alpha\, , \qquad
  W_{\alpha \dot B} {}^i := \frac{i}{7} {\bb D}_{\alpha \dot A} W_{\dot A\dot B}{}^i \, ,
 \ee
 and that
 \be\label{DW=WI}
 {\bb{D}}_{\alpha (\dot A} W_{\dot{B} )\dot C}{}^i  =  i W_{\alpha \dot D}{}^i
 \left(\delta_{\dot D (\dot A} \delta_{\dot B )\dot C} - \delta_{\dot D\dot C}\delta_{\dot A\dot B}\right)\, .
  \ee
 Using the Jacobi identity
 \be
 \left[{\cal D}_\mu, \left[ \bb D_{\beta\dot B}, \bb D_{\gamma \dot C} \right]_+ \right] _-
 + \left[ \bb D_{\gamma \dot C} \left[ \bb D_{\beta\dot B}, {\cal D}_\mu\right]_-\right]_+
 +  \left[ \bb D_{\beta \dot B} \left[ \bb D_{\gamma\dot C}, {\cal D}_\mu\right]_-\right]_+
 \equiv 0\, ,
 \ee
one finds that
\be F_{\mu\nu}{}^i = \frac{1}{8}
\epsilon_{\mu\nu\rho} W^\rho{}^i \, , \qquad W_\mu{}^i :=
\frac{1}{2}  \bar {\bb D}_{\dot A} \gamma_\mu W_{\dot A}{}^i \, ,
\ee
and also that
 \be\label{DW=DW+F}
 \bb{D}_{\dot{A} (\alpha} W_{\beta) \dot{B}}{}^i  =
(C\gamma^\mu)_{\alpha\beta}\left( {\cal D}_\mu
W_{\dot{A}\dot{B}}{}^i - 4 \delta_{\dot{A}\dot{B}} W_\mu
{}^i\right)\, ,  \qquad \bb D_{\alpha (\dot A} W^\alpha_{\dot B)}
{}^i =0\, . \ee We see that the SDiff field strength supermultiplet
includes a scalar  ${\bf 28}$ ($W_{\dot A\dot B}{}^i$), a spinor
${\bf 8}_c$ ($W_{\alpha\dot A}{}^i$) and a singlet divergence-free
vector ($W^\mu{}^i$).  There are many other independent components
but these become dependent on-shell. The  relevant Chern--Simons--like (CS--like)
superfield equation {\it in the absence of `matter'
supermultiplets} is obviously $W_{\dot A\dot B}{}^i=0$, since this
sets to zero all SDiff$_3$ field strengths. We shall see below how
this must be modified in the presence of `matter'.

We now introduce an ${\bf 8}_v$-plet of scalar, and
SDiff$_3$-scalar, superfields $\phi^I$. The lowest component, which
we also call $\phi^I$, may be identified with the BLG scalar fields.
One then expects to find the  superpartners in the next component,
at least on-shell, and they should appear as the lowest component of
an ${\bf 8}_s$-plet  of spinor  superfields $\psi_{\dot A}$.  We
therefore impose the constraint\footnote{This equation  was called
the  {\it superembedding--like} equation  in \cite{Bandos:2008df}
because it  can be obtained from the `superembedding' equation for a
single M2--brane \cite{Bandos:1995zw} by first linearizing with
respect to the dynamical fields in the static gauge, as in
\cite{Howe:2004ib}, and then covariantizing the result with respect
to SDiff$_3$.}
\be\label{DX=rp}
\bb{D}_{\alpha \dot{A}} \phi^I= i \tilde{\rho}^I_{\dot{A}B}
\psi_{\alpha B}\, .
\ee
Acting on this constraint with an  SDiff$_3$-covariant spinor derivative, and
making use of the anticommutation relation (\ref{D21}), one finds
that
\be\label{Daf=W}
\bb{D}_{\alpha [\dot{A} }\tilde{\rho}^{I}{}_{\dot{B}]C}
\psi^\alpha_C = 2 W_{\dot{A}\dot{B}}{}^i\partial_i\phi^I\, , \ee
which is solved by what was called in \cite{Bandos:2008df} the
`super-CS' equation
\be\label{WdAdBi=}
W_{\dot{A}\dot{B}}{}^i=
 {2 g \over {e}}\varepsilon^{ijk}\partial_i\phi^I\partial_j\phi^J\tilde{\rho}^{IJ}_{\dot{A}\dot{B}}\, .
\ee
It was shown in  \cite{Bandos:2008df} that  the two ${\cal N}=8$
superfield equations (\ref{DX=rp}) and (\ref{WdAdBi=}) imply the
Nambu-bracket BLG equations (\ref{EofM}).

\section{Pure-spinor superspace}
\setcounter{equation}{0}

An {\it off-shell} ${\cal N}=8$ superfield formulation of the
abstract BLG theory was proposed by  Cederwall
\cite{Cederwall:2008vd}.  This formulation involves  a `pure-spinor superspace'
for which there is an additional  ${\bf 8}_c$-plet\footnote{Actually,  ${\bf 8}_s$
valued bosonic spinors were used in \cite{Cederwall:2008vd}, but
this is just a matter of convention.}   of  {\it complex} commuting
spinor coordinates $\lambda_{\dot A}$  satisfying the `purity'
condition
\be\label{N8pure}
\bar\lambda \gamma^\mu \lambda =0\, ,
\qquad \left(\bar\lambda := \lambda^t C\right)
\ee
 where the {\it summed} $Spin(8)$ indices have been suppressed.  In other
words, the pure-spinor superspace is parametrized by the standard
${\cal N}=8$ $D=3$ superspace  coordinates $(x^\mu
,\theta^\alpha_{\dot{A}})$ together with $\lambda^\alpha_{\dot A}$.
This is a variant of the $D=10$ pure-spinor superspace first
proposed by Howe \cite{Howe:1991mf} and, from a more general
perspective, a realization of the harmonic superspace
programme of \cite{Galperin:1984av}.  All pure-spinor superfields will be assumed
to be {\it analytic} functions of $\lambda$ that can be expanded as a Taylor
series in powers of $\lambda$. Our aim here is to extend this
formalism to the Nambu bracket realization of the BLG theory in
which all pure-spinor superfields are additionally functions on the
closed 3-manifold $M_3$.

\subsection{Pure spinor Fierz identities}

We begin by establishing some properties of the pure-spinor  $\lambda$. The only analytic nonvanishing pure spinor bilinears are
\be\label{bilinears}
 M_{IJ}:= \bar{\lambda}\, \tilde{\rho}^{IJ}\lambda\, , \qquad
 N^\mu_{IJKL}:= \bar{\lambda}\, \gamma^\mu \tilde{\rho}^{IJKL}\lambda\, .
 \ee
 For example,
 \begin{equation}\label{lAlB=}
\bar{\lambda}_{\dot{A}}\lambda_{\dot{B}} ={1\over 16} M^{IJ}\,
\tilde{\rho}_{\dot{A}\dot{B}}^{IJ}\; , \qquad
\bar{\lambda}_{\dot{A}}\gamma_\mu\lambda_{\dot{B}} = {1\over 16\cdot
4!}\, N_\mu^{IJKL}\, \tilde{\rho}_{\dot{A}\dot{B}}^{IJKL} \, .
\end{equation}

It was stated in  \cite{Cederwall:2008vd} that the constraint (\ref{N8pure}) implies the identity
\be\label{IDone}
M_{IJ} \, \rho^J\lambda  \equiv 0\, .
\ee
This can be proved as follows. A Fierz transformation of the left hand side yields
\begin{eqnarray}\label{FF1}
& M_{IJ}\  \rho^J \lambda = - \frac{1}{8}\, \rho^I\  \left( M_{PQ}\
\rho^{PQ}\ \lambda - \frac{1}{120}N_\mu^{PQRS}\ \gamma^\mu
\rho^{PQRS}\lambda\right)\, ,
\end{eqnarray}
which implies that \be\label{FF1+2} M_{PQ}\,  \rho_{PQ}\lambda =
\frac{1}{120} N^\mu_{PQRS}\, \gamma_\mu \rho_{PQRS}\lambda \, . \ee
A Fierz transformation of the left hand side of {\it this} equation
leads, on using the identities \be \rho_{PQ}\  \rho_{JK}\ \rho_{PQ}
= -8\, \rho_{JK}\, , \qquad \rho_{PQ} \ \rho_{JKLM}\ \rho_{PQ} = 8\,
\rho_{JKLM}\, , \ee to the conclusion that \be\label{A} M_{PQ}\,
\rho_{PQ}\lambda = \frac{1}{72} N^\mu_{PQRS}\, \gamma_\mu
\rho_{PQRS}\lambda \, . \ee Comparing (\ref{A}) with (\ref{FF1+2}),
we see that \be M_{PQ}\, \left(\rho_I \ \rho_{PQ}\right)\lambda =
N^\mu_{PQRS}\, \gamma_\mu \left(\rho_I \rho_{PQRS}\right)\lambda
=0\, , \ee and using this in (\ref{FF1})  we deduce
(\ref{IDone}).

The purity condition on $\lambda$ also implies the following identities, the first of which was
used  in  \cite{Cederwall:2008vd}:
\begin{equation}\label{MMNN}
(a)\ M_{[IJ} M_{KL]}=0 \, , \qquad (b) \ N_{PQ[IJ} \cdot N_{KL]PQ} \equiv 0\, .
\end{equation}
To prove these identities, it is convenient to begin by defining
\begin{equation}\label{MIJ:=}
 M_{IJKLPQ}:= \bar{\lambda}\, \tilde{\rho}^{IJKLPQ}\lambda\; = {1\over 2}
 \epsilon^{IJKLPQRS} M_{RS}  \, ,
\end{equation}
and taking note of the following Spin(8) sigma-matrix identities
\begin{eqnarray}\label{ID1}
\bar\lambda\left( \tilde{\rho}_{[IJ} \ \tilde{\rho}^{PQ} \
\tilde{\rho}_{KL]} \right)\lambda &=& M_{IJKLPQ} + 4 M_{[IJ}
\delta_K{}^P \delta_{L]}{}^Q \nonumber\\
\bar\lambda\gamma_\mu \left(\tilde{\rho}_{[IJ}\ \tilde{\rho}^{PQRS}\
\tilde{\rho}_{KL]} \right)\lambda &=&  24 N_\mu{}^{[RS}{}_{[IJ} \
\delta_K{}^P \delta_{L]}{}^{Q]}\, ,
\end{eqnarray}
\begin{eqnarray}\label{ID2}
\bar\lambda \left(\tilde{\rho}_{IJKL}\ \tilde{\rho}^{PQ}\right)
\lambda &=& M_{IJKLPQ} -12 M_{[IJ}  \
\delta_K{}^P \delta_{L]}{}^Q \nonumber\\
\bar\lambda\gamma^\mu \left(\tilde{\rho}_{IJKL} \
\tilde{\rho}^{PQRS} \right)\lambda &=& -72
 N_\mu{}^{[RS}{}_{[IJ} \
\delta_K{}^P \delta_{L]}{}^{Q]}\, ,
\end{eqnarray}
and
\begin{eqnarray}\label{ID3}
\bar\lambda \left(\tilde{\rho}_{IJKLMN}\ \tilde{\rho}^{PQ}\
\tilde{\rho}_{MN}\right)\lambda &=&
4 \, M_{IJKLPQ} + 144 \, M_{[IJ} \delta_K{}^P\delta_{L]}{}^Q \nonumber\\
\bar\lambda\gamma_\mu \left(\tilde{\rho}_{IJKLMN}\
\tilde{\rho}^{PQRS}\ \tilde{\rho}_{MN}\right) \lambda &=& -288\
N_\mu^{[RS}{}_{[IJ} \delta_K{}^P\delta_{L]}^{Q]}\, .
\end{eqnarray}
Now, performing a Fierz transformation of the left hand side  of
(\ref{MMNN}b),  we deduce, on using (\ref{ID1}), that
\begin{eqnarray}\label{F1}
M_{[IJ} M_{KL]} + \frac{1}{36} M_{RS} M_{IJKLRS} +
\frac{1}{36} N_{PQ[IJ} \cdot N_{KL]PQ} =0 \, .
\end{eqnarray}
Next we note that  the purity condition implies that
\begin{eqnarray}\label{lglN=0}
(\bar\lambda\gamma_\mu \lambda)N^\mu_{IJKL} =0\; . \qquad
\end{eqnarray}
A Fierz transformation of the left hand side leads, on using  (\ref{ID2}), to the equation
\be\label{F2}
 M_{[IJ}M_{KL]}  -  \frac{1}{12}M_{RS}M_{IJKLRS} + \frac{1}{12} N_{PQ[IJ} \cdot N_{KL]PQ} =0\, .
\ee
Finally, a Fierz transformation of $M_{RS}M_{IJKLRS}$, and use of (\ref{ID3}),  leads to the relation
\be\label{F3}
M_{[IJ}M_{KL]}  + \frac{1}{4} M_{RS}M_{IJKLRS}  + \frac{1}{12} N_{PQ[IJ} \cdot N_{KL]PQ} =0\, .
 \ee
One can check that the system of three equations, (\ref{F1}),
(\ref{F2}) and (\ref{F3}) for the three `variables'  $M_{[IJ}
M_{KL]}$, $M_{RS} M_{IJKLRS}$ and $N_{PQ[IJ} \cdot N_{KL]PQ}$, has
only the trivial solution. This proves  (\ref{MMNN}).

\subsection{Off-shell BLG}

Again following \cite{Cederwall:2008vd}, we define the BRST-type
operator
\be Q := \bar \lambda D\, , \ee which
satisfies $Q^2\equiv 0$ as a consequence of the purity condition
(\ref{N8pure}). We also introduce an $M_3$-vector-valued  complex
{\it anticommuting} scalar $\Psi^i$. In the present context,
$\Psi^i$ will play the role of the SDiff$_3$ gauge potential; its
SDiff$_3$ gauge transformation, with commuting $M_3$-vector
parameter $\Xi$, is
 \be\label{QSDiffPi=} \delta \Psi^i
= Q \Xi^i  + \Psi^j\partial_j \, \Xi^i - \Xi^j\partial_j \Psi^i\, ,
\qquad
\partial_i\left(e\Xi^i\right) =0\, .
\ee We require that $\partial_i\left(e\Psi^i\right)=0$ so that,
locally on $M_3$,
 \be\label{Psi=eedPi}
 \Psi^i = e^{-1}
\varepsilon^{ijk}\partial_j \, \Pi_k\, ,
 \ee
where $\Pi_i$ is the  complex {\it anticommuting}, and spacetime
scalar, pre-gauge potential of this formalism. Note that, in
contrast to the rather similar formalism of section
\ref{sec:SDiffGI},  the gauge potential and pre-potential are
Minkowski  {\it scalars} (albeit anticommuting)  rather than
one-forms\footnote{This is not so surprising when one recalls that
the exterior product of  `bosonic'  one-forms provides a representation of
Grassmann algebra multiplication.}.

Next, following our ${\cal N}=1$ superspace discussion at the end
of  subsection \ref{subsec:superspace}, we may introduce the
field-strength superfield
 \be\label{FSS}
 {\cal F}^i := Q \Psi^i + \Psi^j
\partial_j \Psi^i\,  = e^{-1}\varepsilon^{ijk}\partial_j {\cal
G}_k\, ,
\ee
where the last equality is valid locally on $M_3$ and
\be
{\cal G}_i  := Q\Pi_i + \Psi^j \partial_j \Psi_i\,
\ee
is the pre-field-strength superfield of this formalism. Both ${\cal
F}^i$ and ${\cal G}_i$ are SDiff$_3$ covariant, so ${\cal F}^i{\cal
G}_i$ is an SDiff$_3$ scalar and its integral is also pre-gauge
invariant (i.e. invariant under $\delta\Pi_i=
\partial_i\alpha$ with an arbitrary anticommuting scalar $\alpha$).
Furthermore, this integral is $Q$-exact, in the sense that
\be
 \int d^3\sigma \, e\, {\cal F}^i  {\cal G}_i = Q \, \bb L_{CS}\, ,
\ee
where
\be
\bb L_{CS} = \int d^3\sigma \, e\,  \left( \Pi_i \, Q\Psi^i -
\frac{1}{3}\epsilon_{ijk}\Psi^i\Psi^j\Psi^k \right)
\ee is the CS-type Lagrangian density of this formalism; it is the Nambu-bracket version
of  the term  proposed in \cite{Cederwall:2008vd} for the abstract BLG theory,
although our construction is different. Note that $\bb L_{CS}$
is both complex and anti-commuting.

We now introduce  the ${\bf 8}_v$-plet of  complex scalar ${\cal
N}=8$ `matter' superfields $\Phi^I$, with SDiff$_3$ variation
 \be\label{dPhi=Xid}
 \delta\Phi^I = \Xi^i\partial_i \Phi^I\, .
 \ee
We allow these superfields  to be complex because they may depend on
the complex  pure-spinor $\lambda$  but,  to make contact with the
on-shell ${\cal N}=8$ superfield equations of  subsection (\ref{subsec:on-shell}),
we will need to impose a reality condition such  that
\be
 \Phi^I = \phi^I + {\cal O}\left(\lambda\right) \, ,
 \ee
where $\phi^I$ is a {\it real} ${\bf 8}_v$-plet of  `standard'  ${\cal N}=8$
scalar superfields.  We also define an SDiff$_3$-covariant extension
of $Q\Phi^I$ by
 \be \label{bbQ=Q+}
 \bb Q \Phi^I := Q\Phi^I + \Psi^i\partial_i \Phi^I \, .
 \ee
We must use this SDiff$_3$-covariant quantity to construct a
`matter' Lagrangian that can be added to the `CS' term, which means
that it must also be anti-commuting and analytic in $\lambda$.  One
possibility is
 \be
{\bb L}_{mat} = {1\over 2}M_{IJ} \oint d^3\sigma \, e \Phi^I \bb Q
\Phi^J\, ,
\ee
with $M_{IJ}$ as defined in
(\ref{bilinears}). To ensure manifest ${\cal N}=8$ supersymmetry one
still needs to specify an adequate superspace integration measure.
We refer to \cite{Cederwall:2008xu} for details of this measure,
which has the crucial property of allowing us to discard a
BRST-exact terms when varying with respect $\Phi^I$. This variation
yields the superfield equation \be\label{mattereq} M_{IJ} \bb
Q\Phi^J =0\, ,
 \ee
 which implies, as a consequence of the identity
(\ref{MMNN}a), that
 \be
 Q\Phi^I = \bar\lambda \tilde \rho^I \Theta
 \ee
for some ${\bf 8}_s$-plet of complex spinor superfields
$\Theta_{\alpha A}$. The first nontrivial  ($\sim\lambda$)  term in the $\lambda$-expansion of this equation is   precisely the on-shell superspace constraint (\ref{DX=rp}) with
$\psi=\Theta|_{\lambda=0}$, which is real as a consequence of the
assumed reality of $\phi^I$.

The combined  SDiff$_3$-invariant, complex and anti-commuting,
Lagrangian density
 \be\label{matter+GF}
 {\bb L} = {\bb L}_{mat} - \frac{1}{g}{\bb L}_{CS}
\ee
is therefore a candidate for an off-shell ${\cal N}=8$ superfield formulation of the Nambu-bracket
realization of the BLG theory, along the lines of \cite{Cederwall:2008vd}.  The $\Pi_i$ equation of motion of this combined  Lagrangian is
\be\label{calFeq}
{\cal F}^i =  \frac{g}{2e}  M_{IJ}\epsilon^{ijk}\partial_j\Phi^I\partial_k\Phi^J \, .
\ee
At this stage it is important to assume that $\Psi^i$  has `ghost number one'  \cite{Cederwall:2008vd}, which means that  it is a power series in $\lambda$ with vanishing zeroth order term
(and similarly for its pre-potential $\Pi_i$). In other words
\be\label{Psi=lsigma}
\Psi^i = \lambda^\alpha_{\dot{A}} \varsigma^i_{\alpha\dot{A}}\; ,
 \ee
where $\varsigma^i$ is an $M_3$-vector-valued  ${\bf 8}_c$-plet of
arbitrary anticommuting spinors. Its zeroth component  in the
$\lambda$-expansion is the fermionic  SDiff$_3$ potential
introduced, with the {\it same symbol}, in (\ref{bbDs}).  With this
`ghost number' assumption, (\ref{calFeq}) produces  at lowest
nontrivial order ($\sim \lambda^2$) the superspace constraints
(\ref{D21}) for the  `ghost number zero' contribution
$\varsigma^i\vert_{\lambda=0}$ to the pure spinor superfield
$\varsigma^i$ in (\ref{Psi=lsigma}), accompanied by the super CS
equation (\ref{WdAdBi=}) for the field strength $W_{\dot{A}\dot{B}}$
constructed from this potential.

We have now shown how  the on-shell ${\cal N}=8$ superfield formulation of subsection \ref{subsec:on-shell},  and hence all  BLG field equations, may be extracted from the equations of motion derived from  the pure spinor superspace action (\ref{matter+GF}).
Of course, the field content and equations of motion should  be
analyzed at all higher-orders in the $\lambda$-expansion.
Our results are consistent with the  conjecture that the field equations
of the action (\ref{matter+GF}) are equivalent to those of the on shell superfield
formulation of  \ref{subsec:on-shell}, in which case our results would
imply  that all higher-order fields in the  $\lambda$ expansion are auxiliary.
Our results are also consistent with the weaker conjecture that all `higher-order' fields are either auxiliary or decouple, in which case they might be removed by some ghost-number
constraint. We shall not attempt to prove either of these conjectures here. Instead,
we limit ourselves to the observation that  a full analysis must take into
account the existence of additional gauge invariances
\cite{Cederwall:2008vd,Cederwall:2008xu}; in the present context,
one may use the identities (\ref{MMNN}) to show that the BLG action
is invariant under the infinitesimal transformations
\begin{eqnarray}\label{QgaugePi=}
\delta \Phi^I= \bar{\lambda}\tilde{\rho}^I \zeta_\alpha + (\bb{Q}+
{\Psi}^j\; \partial_j) K^I\; ,  \qquad \delta {\Pi}_i = K^I\,
M_{IJ}\,
\partial_i\Phi^J\; ,
\end{eqnarray}
for arbitrary pure-spinor-superfield parameters $\zeta_\alpha$ and
$K^I\;$.

\section{Discussion}
\setcounter{equation}{0}

It has been known for some time that there exist  Yang-Mills gauge theories, in $D$-dimensional  Minkowski  spacetime, for which the gauge group is the infinite-dimensional group of area-preserving diffeomorphisms SDiff($M_2$) of  $M_2$, a closed  two-dimensional manifold  that is compact with respect to some volume form.  The manifold $M_2$ plays the role of an `internal' space on which
all Minkowski-space fields are also tensors, e.g. functions.  Such models first arose for $D=1$ as gauge-mechanics models governing the light-cone-gauge dynamics of a relativistic membrane  \cite{LCGM2-SU(inf),de Wit:1988ig}; it was later appreciated that the construction applies for any $D$ \cite{Floratos:1988mh}.  A natural question is whether there exist gauge theories for which the gauge group is the group  SDiff($M_n$) of volume-preserving diffeomorphisms of some $n$-dimensional manifold $M_n$ for $n\ge 3$; we assume that $M_n$ is closed and compact with respect to some volume $n$-form.  Examples, with $n=p$, may be found for $D=1$ by light-cone gauge fixing of a relativistic  p-brane \cite{Bergshoeff:1988hw}, but no gauge-field kinetic term is required in this  case.  In this paper, we have developed a general formalism for the construction of  $D>1$  gauge theories of $n$-volume-preserving diffeomorphisms. We ignored some global issues that distinguish between manifolds $M_n$ of different topology, partly because we are mostly interested in  the simplest  case in which $M_n$ is the $n$-sphere; for that reason we abbreviated SDiff($M_n$) to SDiff$_n$.

The construction of a gauge-field kinetic term  for an SDiff$_n$ gauge theory is obstructed by the absence of  a metric on $M_n$ (as any metric could not be SDiff$_n$ inert, it would have to be introduced as a dynamical variable and then we would have some GR-type theory rather than a Minkowski field theory).  As far as we can see, this obstacle is
insuperable for $n\ge4$,  but there are options for $n=3$.  In
particular, we have constructed a SDiff$_3$ invariant analog of the
$D=10$ super-YM theory.  This theory is unphysical because the
energy is not positive definite but it is nevertheless an example of
an `exotic' $D>3$ Minkowski-space gauge theory; i.e. one not of YM
type.  This shows that the uniqueness of the YM minimal interaction
for $D>3$ \cite{Henneaux:1997bm,Barnich:2000zw} fails to apply when
the number of massless  vector fields is infinite. For $D=3$ there
is another possibility for the construction of an SDiff$_3$
invariant gauge-field  kinetic term; this is an analog of the YM
Chern-Simons (CS) term although the SDiff$_3$ version is parity even
because a parity flip in Minkowski spacetime can be `undone'  by a
parity flip in the `internal' 3-space. We  have shown how to
construct a general class of ${\cal N}=1$ supersymmetric SDiff$_3$
gauge theories with this CS-type kinetic term, in components and
using superspace methods.

Of particular interest is the  special case of the superconformal
$D=3$ SDiff$_3$ gauge theory with maximal ${\cal N}=8$
supersymmetry, because this is the Nambu-bracket realization of the
BLG  theory \cite{Bagger:2007jr,Gustavsson:2007vu}, which can be
viewed as  describing a `condensate' of coincident planar M2-branes;
this realization  was first  considered by Bagger and Lambert
\cite{Bagger:2007vi}, but  the CS-type term appears first in
\cite{Ho:2008ve}.  We have presented here the full Lagrangian and
supersymmetry transformation laws in a simple form.
Following the original version of this paper, an  ${\cal N}=8$ superspace
formulation of the SDiff$_3$ gauge theory was proposed  by one of us
\cite {Bandos:2008df}, and we have reviewed this work, presenting
some additional simplifications. This formalism makes the ${\cal
N}=8$ supersymmetry manifest, although only at the level of the
equations of motion. An alternative off-shell ${\cal N}=8$
superfield formalism of the abstract BLG theory was proposed  around
the same time by Cederwall \cite{Cederwall:2008vd,Cederwall:2008xu};
his formalism uses fields defined on a pure-spinor extension  of
${\cal N}=8$ superspace. We have shown here how this pure-spinor
superspace formalism can be fused with our  SDiff$_3$ formalism to
give an off-shell action for the M2 condensate, although we did not attempt
a full analysis of the field content.

The BLG theory was found by requiring that the Basu-Harvey equation \cite{Basu:2004ed},
proposed to describe $N$ M2-branes ending on an M5-brane, should arise as a condition
for preservation of 1/2 supersymmetry. The original  equation is solved by a tube-like
configuration with  a `fuzzy' 3-sphere cross-section but this fuzzy 3-sphere becomes
a smooth 3-sphere in the Nambu-bracket realization \cite{Bagger:2007vi}. Here we have verified that this  `smoothed' Basu-Harvey equation is an  equation for preservation of 1/2 supersymmetry
in the context of the SDiff$_3$ invariant theory for an M2 condensate. This could be viewed
as further evidence of the connection between the BLG theory and the M5-brane   \cite{Ho:2008nn,Ho:2008ve} although we believe this connection has not yet been properly understood; our current views on this topic can be found in  \cite{Bandos:2008fr}.

In the special case that $M_3=M_2\times S^1$, we have shown that one may perform
a  dimensional reduction of the SDiff$_3$ invariant  BLG theory to arrive at an SDiff$_2$-invariant $D=3$ Yang-Mills gauge  theory with maximal supersymmetry, which we interpreted as a model governing the low-energy dynamics of  a  D2-brane condensate of IIA superstring theory; recall  that   SDiff$_2$ may be loosely viewed as the $N\to\infty$ limit of $SU(N)$, and that
the low-energy dynamics of a  collection of $N$ planar  D2-branes is governed by a maximally supersymmetric $D=3$ $SU(N)$ gauge theory.  Results of  \cite{Shimada:2003ks}
suggest that different ways of taking the large $N$  limit of $SU(N)$ lead to
different topologies for $M_2$, and we imagine that something similar might  apply to
$M_3$ in the case of the M2-brane condensate. This issue is connected to the
important question that we passed over in the introduction: the nature of the low-energy dynamics
of $N$ coincident planar M2-branes for {\it finite} $N$.

It is tempting to suppose  that  an action describing the infra-red dynamics of  $N$ coincident M2-branes can be obtaned  by some discretization of the Nambu-bracket 3-algebra of functions on $S^3$, but this idea runs into the difficulty, mentioned the introduction, that there is no suitable sequence of finite-dimensional metric Filippov 3-algebras labelled by $N$. There have been several proposals to circumvent this difficulty. One is to consider other types of algebra, e.g. \cite{Cherkis:2008qr}.
Another is  to allow
non-metric Filippov 3-algebras, which means  that one is restricted
to consider equations of motion; in this scheme there is a natural
explanation for the expected $N^{3/2}$ scaling of the number of
degrees of freedom with the number $N$ of M2-branes
\cite{Chu:2008qv} (see also \cite{Berman:2006eu}). Basically, fields
on $S^3$ become $n\times n\times n$ `cubic matrices'  with $\sim
n^3$ degrees of freedom. However, the potential vanishes for fields
on $S^3$ that depend on only two of its coordinates, and these
become `standard' $n\times n$ matrices with $\sim n^2$ degrees of
freedom. The moduli space of vacua therefore has dimension $\sim
n^2$, so that the number $N$ of M2-branes described by the model
scales with $n$ like $n^2$; the number of degrees of freedom
therefore scales with $N$ like $N^{3/2}$, exactly as predicted by
AdS/CFT \cite{Klebanov:1996un}.

This `success' of the
Nambu-bracket approach may be contrasted with  currently
popular `ABJM' proposal that involves an  $U(N)\times U(N)$ CS theory at
level $k=1$, with bi-fundamental matter \cite{Aharony:2008ug}; this
model has a manifest ${\cal N}=6$ supersymmetry but is conjectured
to be ${\cal N}=8$ supersymmetric. It is a `conventional' theory in
the sense that its construction does not involve 3-algebras, but it
is strongly coupled and so one cannot expect to read off the degrees
of freedom from the Lagrangian.  This is just as well since the
conventional gauge theory structure would lead one to expect the
number of degrees of freedom to scale like $N^2$, so one is led to
conjecture that this is reduced to $N^{3/2}$ by strong coupling
effects. Although there is considerable support for this proposal,
e.g. \cite{Berenstein:2008dc,Hosomichi:2008ip,Drukker:2008jm}, it seems to us
that it is more like a restatement of the problem (to one of strong coupling dynamics)
than a solution to it. 

If the ABJM proposal is correct, as seems likely, it should be possible to take the limit of 
large $N$ to find the theory describing the M2-condensate, which could then be compared
with the SDiff$_3$ gauge theory  presented in detail here. However, this
would involve taking {\it two} limits simultaneously, strong coupling and large $N$. Double limits 
are notoriously tricky; they may not commute. It seems quite possible that one such limit
could yield the ${\cal N}=8$ supersymmetric SDiff$_3$ gauge theory, so there is no logical 
contradiction between  the Nambu bracket approach advocated here and the conventional CS
approach of ABJM.

Another outstanding problem is the nature of the $D=6$ conformal
field theory governing the low energy dynamics of $N$ coincident
M5-branes.  In light of what we now know about multiple coincident
M2-branes, it seems likely that this problem will simplify in the
$N\to\infty$ limit. Given that  a condensate of M2-branes may be viewed, in some
sense, as an M5-brane, then is there a similar sense in which an M5 condensate
could be viewed as a yet higher-dimensional M-brane? Recalling that
the recent advances in the  M2 case were prompted by the Basu-Harvey
proposal that  the boundary of multiple M2-branes on an M5-brane
might be understood in terms of fuzzy 3-spheres, it is natural to
reconsider the implications of the recent demonstration
\cite{Bergshoeff:2006bs} that an M5-brane can have a boundary on an
M9-brane, which is a boundary of the 11-dimensional bulk spacetime
of M-theory; in this context we should mention that
higher-dimensional generalizations of the Basu-Harvey equation have
been considered in \cite{Berman:2006eu,Bonelli:2008kh}.

\section*{Acknowledgments}

IAB thanks Dmitri Sorokin, Paolo Pasti and Mario Tonin for
the hospitality in Padova where part of this work was done. IAB
is supported by the Basque Science Foundation {\it
Ikerbasque} and partially by research grants from the Spanish MCI
(FIS2008-1980), the INTAS (2006-7928), and the Ukrainian National
Academy of Sciences and Russian RFFI grant 38/50--2008. PKT is
supported by an EPSRC Senior Research Fellowship, and he thanks the
University of Barcelona for hospitality.


\end{document}